\documentclass[aps,prx,twocolumn,superscriptaddress,longbibliography]{revtex4-1}

\usepackage{amssymb}
\usepackage{amsmath}
\usepackage{graphicx}
\usepackage{amsfonts}
\usepackage{braket}
\usepackage{subfigure}
\usepackage{hyperref}
\usepackage[version=3]{mhchem}

\begin{document}
\title{Dynamical  birefringence: Electron-hole recollisions as probes of Berry curvature}
\author{Hunter B. Banks}
\affiliation{Physics Department, University of California, Santa Barbara, USA}
\affiliation{Institute for Terahertz Science and Technology, University of California, Santa Barbara, USA}
\author{Qile Wu}
\affiliation{Department of Physics, The Chinese University of Hong Kong, Hong Kong, China}
\affiliation{Department of Physics, Tsinghua University, Beijing 100084, People's Republic of China}
\author{Darren C. Valovcin}
\affiliation{Physics Department, University of California, Santa Barbara, USA}
\affiliation{Institute for Terahertz Science and Technology, University of California, Santa Barbara, USA}
\author{Shawn Mack}
\affiliation{U.S. Naval Research Laboratory, Washington, DC, USA}
\author{Arthur C. Gossard}
\affiliation{Materials Department, University of California, Santa Barbara, USA}
\author{Loren Pfeiffer}
\affiliation{Electrical Engineering Department, Princeton University, Princeton, NJ, USA}
\author{Ren-Bao Liu}
\affiliation{Department of Physics, The Chinese University of Hong Kong, Hong Kong, China}
\author{Mark S. Sherwin}
\affiliation{Physics Department, University of California, Santa Barbara, USA}
\affiliation{Institute for Terahertz Science and Technology, University of California, Santa Barbara, USA}

\date{\today}

\begin{abstract}
The direct measurement of Berry phases is still a great challenge in condensed matter systems. The bottleneck has been the ability to adiabatically drive an electron coherently across a large portion of the Brillouin zone in a solid where the scattering is strong and complicated. We break through this bottleneck and show that high-order sideband generation (HSG) in semiconductors is intimately affected by Berry phases. Electron-hole recollisions and HSG occur when a near-band gap laser beam excites a semiconductor that is driven by sufficiently strong terahertz (THz)-frequency electric fields. We carried out experimental and theoretical studies of HSG from three GaAs/AlGaAs quantum wells. The observed HSG spectra contain sidebands up to the 90th order, to our knowledge the highest-order optical nonlinearity reported in solids. The highest-order sidebands are associated with electron-hole pairs driven coherently across roughly 10\% of the Brillouin zone around the $\Gamma$ point. The principal experimental claim is a dynamical birefringence: the intensity and polarization of the sidebands depend on the relative polarization of the exciting near-infrared (NIR) and the THz electric fields, as well as on the relative orientation of the laser fields with the crystal. We explain dynamical birefringence by generalizing the three-step model for high-order harmonic generation. The hole accumulates Berry phases due to variation of its internal state as the quasi-momentum changes under the THz field. Dynamical birefringence arises from quantum interference between time-reversed pairs of electron-hole recollision pathways.  We propose a method to use dynamical birefringence to measure Berry curvature in solids.
\end{abstract}
\maketitle

\section{Introduction}
  When parameters in a quantum system change adiabatically, the quantum states of the system accumulate both dynamic and Berry phases~\cite{berry1984quantal}. Dynamic phases are associated with the energy eigenvalues of the system, while Berry phases are associated with adiabatic changes in wave functions of the system through a quantity called the Berry curvature. Berry phases are of fundamental importance in many branches of physics, such as quantum field theories~\cite{senthil2000}, optics~\cite{Kwiat1991}, ultra-cold atoms~\cite{lin2009}, quantum computing~\cite{neeley2009}, and condensed matter physics~\cite{Chang2008,Xiao2010}. In condensed matter systems, Berry phases are accumulated when a Bloch electron moves along a path in quasi-momentum space~\cite{Chang2008,Xiao2010}. Many manifestations of Berry phases in condensed matter physics have been observed, such as quantum Hall effects, anomalous Hall effects, and Faraday rotations~\cite{Kohmoto1993,Murakami2003,Yang2014}. 
  In materials that exhibit these and related phenomena, the Berry curvature of an energy band is as important as its dispersion relation. However, although there has been recent progress in ultra-cold atoms~\cite{Zak1989,Atala2013,Flaschner1091} and optical systems~\cite{Wimmer2017}, Berry curvature has largely resisted direct experimental measurement in solids~\cite{Virk2011,Priyadarshi2015} because it is difficult to coherently and adiabatically drive an electron across a large portion of the Brillouin zone without the quantum pathway being destroyed by scattering.
	
  Strong laser fields can accelerate Bloch electrons across a substantial fraction of the Brillouin zone in times shorter than typical scattering times, and hence provide opportunities to probe Berry curvature.  In solids, high-order harmonic generation (HHG) results when laser fields in the V/\AA\ (100 MV/cm) range at wavelengths longer than 3 $\mu$m (photon energy less than $0.4$ eV) create electrons and holes by Zener tunneling across the band gap, and subsequently drive them across a substantial fraction of the Brillouin zone~\cite{Ghimire2010,Schubert2014,Hohenleutner2015,Vampa2015nat,chin2001extreme}.  It has been suggested that HHG can be used for an all-optical reconstruction of band structures~\cite{Vampa2015prl}, and that Berry phases may play a role in the HHG process~\cite{Ghimire2010,Liu2017}. However, the combination of diabatic Zener tunneling events with subsequent adiabatic accelerations of electrons and holes within bands complicates the theoretical modeling of HHG~\cite{Langer2016}, and has led to a range of theoretical approaches~\cite{Ghimire2010, Vampa2014,Vampa2015nat, Vampa2015prb,wu2015high,Higuchi2014,Hawkins2015}.  
  
  High-order sideband generation (HSG)~\cite{Liu2007, Zaks2012} is a process similar to HHG, in which the strong laser field that accelerates Bloch electrons is separate from the weak laser that creates them.  HSG can be described by a three-step model similar to that originally proposed by Corkum in the context of HHG in atoms~\cite{Corkum1993}. Electron-hole pairs are created in a controllable and well-defined initial state  by a weak near-infrared (NIR) laser tuned near the  band gap of a semiconductor. The semiconductor is simultaneously driven by a strong terahertz (THz)-frequency electric field, which is strong enough to accelerate the electron and hole into large-amplitude, coherent trajectories in momentum space, but not strong enough to create electron-hole pairs by Zener tunneling.  After their adiabatic evolution in momentum space, electron-hole pairs that recollide and recombine emit sidebands with higher photon energies than the NIR laser that created them.  HSG was predicted by one of us~\cite{Liu2007}, and has been observed in GaAs quantum wells (QWs)~\cite{Zaks2012,Banks2013}, bulk GaAs~\cite{Zaks2013}, and, in time-resolved experiments, in bulk \ce{WSe2}~\cite{Langer2016}.  Recent theoretical work has pointed to the importance of Berry curvature to HSG~\cite{Yang2013,Yang2014a,Yang2014}. The Berry phases arise due to variation of the internal states of the electron and hole with varying quasi-momentum in the Brillouin zone. The polarizations of high-order sidebands are affected by quantum interferences between time-reversed pairs of quantum trajectories that have opposite Berry phases~\cite{Yang2013,Yang2014a,Yang2014}.

  Previous measurements of HSG showed that low-order sidebands from InGaAs QWs were slightly weaker when the electric field of the NIR laser was polarized perpendicular to the electric field of the THz beam than when the fields were parallel~\cite{Zaks2012}. This observation is surprising: within the three-step model, and given the cubic symmetry of GaAs, why should the polarization of the NIR laser that created the electron-hole pairs affect the intensity of the sidebands caused by their recollision? A subsequent theoretical investigation predicted that in bulk GaAs the highest-order sidebands generated when NIR and THz laser polarizations were perpendicular should be stronger than when they were parallel~\cite{Liu2015}.
	
  In this paper, we carried out systematic experimental and theoretical studies of HSG from three GaAs/AlGaAs quantum wells driven by 40 ns pulses of linearly-polarized 540 GHz radiation with a strength of 35 kV/cm in the quantum wells. The observed HSG spectra contained sidebands up to the 90th order, to our knowledge the highest-order optical nonlinearity reported in solids. The highest-order sidebands were associated with electron-hole pairs driven coherently across roughly 10\% of the Brillouin zone around the $\Gamma$ point, making the Berry phase effects especially relevant. Although GaAs exhibits neither birefringence (polarization-dependent refraction) nor dichroism (polarization-dependent absorption), we observed surprising polarization-dependent effects in HSG that we call ``dynamical birefringence":  at sufficiently high orders, the sideband \textit{intensities} are usually larger when the exciting NIR and the THz electric fields are polarized perpendicular than when they are parallel, and also depend on the angle between the THz electric field and the crystal axes of the GaAs quantum wells; and sideband \textit{polarizations} exhibit significant ellipticity that increases with increasing order even though the polarizations of both the exciting NIR laser and the THz field are nearly linear.  
  
  To understand dynamical birefringence we generalize the three-step model for HHG to the case of HSG, including both the effects of band structure and Berry curvature. The hole accumulates Berry phases due to variation of its internal state as the quasi-momentum changes under the THz field. Dynamical birefringence arises from quantum interference between time-reversed pairs of electron-hole recollision pathways, which are associated with different Berry phases. The observation and theoretical understanding of the dynamical birefringence in HSG open the door to direct measurements of complete electronic structures of semiconductors and insulators near the $\Gamma$ point, including band structures, scattering rates, and Berry curvatures.

\section{Experimental Results} \label{ExpResults}
  High-order sideband generation (HSG) experiments were performed on three samples with different degrees of quantum confinement and disorder. All the samples contained multiple \ce{Al_{x}Ga_{1-x}As} quantum wells (QWs) separated by \ce{Al_{0.3}Ga_{0.7}As} barriers grown on (100) GaAs substrates~\footnote{To denote lattice directions and planes, we use the convention where $[hkl]$ denotes a direction in the lattice and $(hkl)$ denotes the plane perpendicular to that direction, where we have chosen $h, k, l = x, y, z$. Later, we will rotate the sample about the [100] axis.}. The sample with the strongest quantum confinement contained twenty 5 nm GaAs QWs. Fluctuations in the widths of such narrow quantum wells cause fluctuations in the 2D band gap that manifest themselves in the widths of the excitonic absorption peaks~\cite{christen1991} (see Appendix~\ref{appAbsorption} for the absorption spectra from all three samples). The second sample contained twenty 10 nm \ce{Al_{0.05}Ga_{0.95}As} QWs. In this sample, well-width fluctuations are smaller than for the 5 nm QWs, but alloy disorder due to local fluctuations in the concentration of aluminum atoms in the well region causes fluctuations in the 2D band gap, which are manifested in the significant broadenings of the excitonic absorption peaks of this sample. The third sample was the least disordered, and contained ten 10 nm GaAs QWs that were grown with special care to ensure the smoothest possible sidewalls, and hence excitonic absorption peaks much narrower than the other two samples. 
    
  To generate high-order sidebands, a continuous-wave NIR laser was tuned just above the frequency of the lowest exciton absorption line (heavy hole) while the samples were driven by 40 ns pulses of 540 GHz radiation from the UC Santa Barbara MM-Wave Free-Electron Laser. The THz electric field in the QWs was $35\pm2$ kV/cm for all experiments. The samples were held at a temperature of 15 K. A schematic of the HSG experiment is shown in Fig.~\ref{schematic}(a). Details about the samples and the experimental methods are in Supplementary Material~\footnote{\label{si_growth}See ``Experiment'' in Supplementary Materials for details on samples and preparation, and the experimental setup.}.
    
  \begin{figure}
    \includegraphics[width=\linewidth]{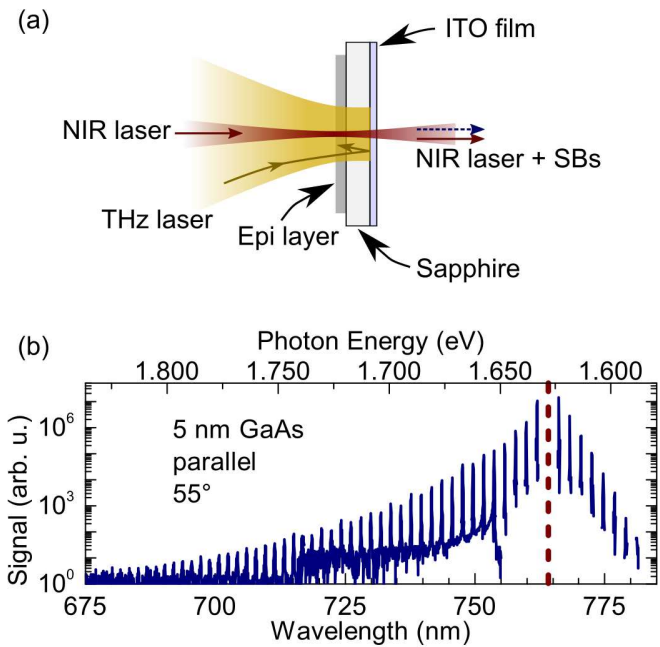}
    \caption{Optical setup and HSG spectrum. (a) Details of the sample. The sample consists of epitaxially grown quantum wells transferred to a sapphire wafer. A thin film of ITO on the back side of the sapphire reflects the THz light back towards the quantum wells while transmitting radiation with NIR wavelengths. See Supplementary Material~\cite{Note2} for more details. For sideband measurements, the NIR and THz lasers are focused collinearly on the same spot on the sample, and propagate normal to the surface. (b) Full HSG spectrum spanning 106 orders from the 5 nm GaAs sample. The sidebands (solid lines) decorate the NIR laser represented by thick dashed red line. The NIR and THz laser polarizations are parallel to each other, and the [011] direction of the lattice makes a 55$^\circ$ angle with the THz polarization. 
    }	
    \label{schematic}
  \end{figure}
	
  \subsection{High-order sideband generation}  
  During simultaneous THz and NIR illumination, the NIR radiation transmitted through the quantum wells contained dozens of sidebands at sideband frequencies $f_\text{SB} = f_\text{NIR} + n f_\text{THz}$, where $f_{\text{NIR}}$ is the frequency of the NIR laser, $f_{\text{THz}}$ is the frequency of the THz field, and $n$ is the sideband order. The HSG spectra for the three samples were all similar to the spectrum shown in Fig.~\ref{schematic}(b). Because these samples were grown on (100) GaAs, a plane with inversion symmetry, only sidebands with even $n$ were observed. At wavelengths longer than the NIR laser line, sidebands with $n \leq -2$ in Fig.~\ref{schematic} (b) decay exponentially with $n$. These sidebands are associated with perturbative nonlinear optical processes ~\cite{Wagner2009,Wagner2011}. At wavelengths shorter than the NIR laser line, sidebands in Fig.~\ref{schematic}(b) are visible with $n$ up to 90, more than three times the highest order previously observed in experiments done on the same sample~\cite{Banks2013}. The increase in the number of observable sidebands is due to a dramatically improved detection scheme and a stronger THz field. All three samples showed sidebands with order up to at least $60$, and each spectrum spanned more than 150 meV.  
        
  The large number of sidebands observed in the HSG spectra reported here enable systematic testing of a three-step model of high-order sideband generation. In such a model, each sideband is associated with an electron-hole pair that recollides and recombines after acceleration through the band structure of a quantum well. Thus, as sideband order increases, so does the fraction of the Brillouin zone explored by the electron and hole, enhancing sensitivity to non-parabolic features of the band structure and to mixing between subbands. 
          
  \subsection{Dynamical birefringence}
\begin{figure}
 	\includegraphics[width=\columnwidth]{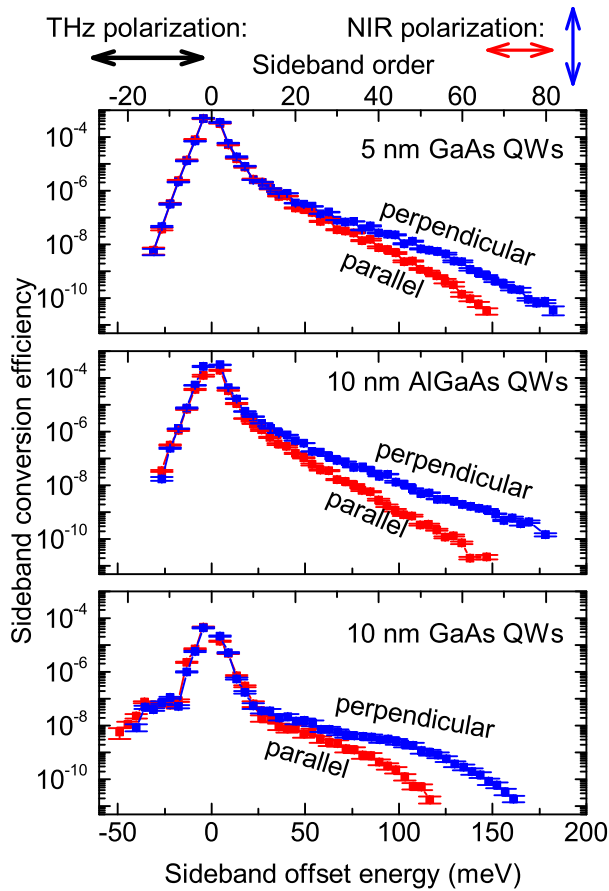}
	\caption{Sideband conversion efficiencies for NIR laser field polarized parallel and perpendicular to THz field. The angles between the THz polarizations and the [011] direction are 85$^\circ$, 91$^\circ$, and 93$^\circ$ for the 5 nm GaAs sample, the 10 nm AlGaAs sample and the 10 nm GaAs sample respectively. The sideband offset energy is the difference of the sideband photon energy and the NIR laser photon energy. The sideband conversion efficiency is the power in the sideband divided by the power of the NIR laser incident on the sample. The perpendicular-excited sidebands are stronger than the parallel-excited sidebands above an energy offset of roughly 70 meV for the 5 nm GaAs sample and 30 meV for the two 10 nm samples. The perpendicularly-excited sidebands fall off more slowly above this offset than below it. In both the 5 nm GaAs and 10 nm AlGaAs samples, the sideband intensities fall off almost exponentially as the sideband order increases. Sideband intensities from the 10 nm GaAs sample, however, show a more complicated relationship with the sideband order, perhaps because of weaker scattering in this sample.  The non-exponential decay of negative-order sidebands in this sample may be due to a relatively thick GaAs layer outside of the QWs (see Supplementary Material~\cite{Note2}).}
	\label{pol_comp}
  \end{figure}

  In all three samples, the intensities of sidebands depend strongly on the relative polarization of the NIR and THz lasers at sufficiently high sideband offset energy (or order). The sideband offset energy is the sum of the kinetic energy of the electron and hole at recollision and a 5-10 meV detuning of the NIR laser below the 2D band gap (see Supplementary Material~\cite{Note2}). In the 5 nm GaAs sample (Fig.~\ref{pol_comp}, top panel), the onset of polarization dependence was at about 70 meV ($n \approx 30$). Sidebands with offset energies above 70 meV were stronger when the NIR laser field was polarized perpendicular to the THz field than when the fields were parallel.  Positive-order sidebands with offset energies below 70 meV, and negative-order sidebands, had intensities that were not noticeably dependent on the NIR laser polarization.  The 10 nm AlGaAs and GaAs samples exhibited similar behaviors, starting at an offset of about 30 meV ($n \approx 16$) (Fig.~\ref{pol_comp}, middle and lower panels). 
  
  The dependence of sideband intensity on the NIR laser polarization is reminiscent of the linear optics of uniaxial crystals, which exhibit birefringence if the refraction of light (\textit{i.e.}, the real part of the index of refraction) depends on the angle of electric field of light with respect to an optical axis, and dichroism if the absorption of light (\textit{i.e.}, the imaginary part of the index of refraction) depends on said angle~\cite{obbornwolf}.  However, GaAs, having a cubic lattice, has no optical axis and is not birefringent or dichroic. As we will show, the strong THz field defines a dynamical optical axis, and the polarization response shown in Fig.~\ref{pol_comp} is a manifestation of what we call ``dynamical birefringence''. For simplicity, we define dynamical birefringence to encompass both effects analogous to refraction (which primarily affect the relative phases of x- and y-components of a sideband's electric field, and hence the sideband's polarization) and effects analogous to absorption or emission (which primarily affect the relative amplitudes of sidebands excited with different polarizations that is shown in Fig.~\ref{pol_comp}).  The correlation between the onset of dynamical birefringence and the well width suggests that quantum confinement effects on the band structure influence this phenomenon.  

\begin{figure*}
		\includegraphics[width=1.0\textwidth]{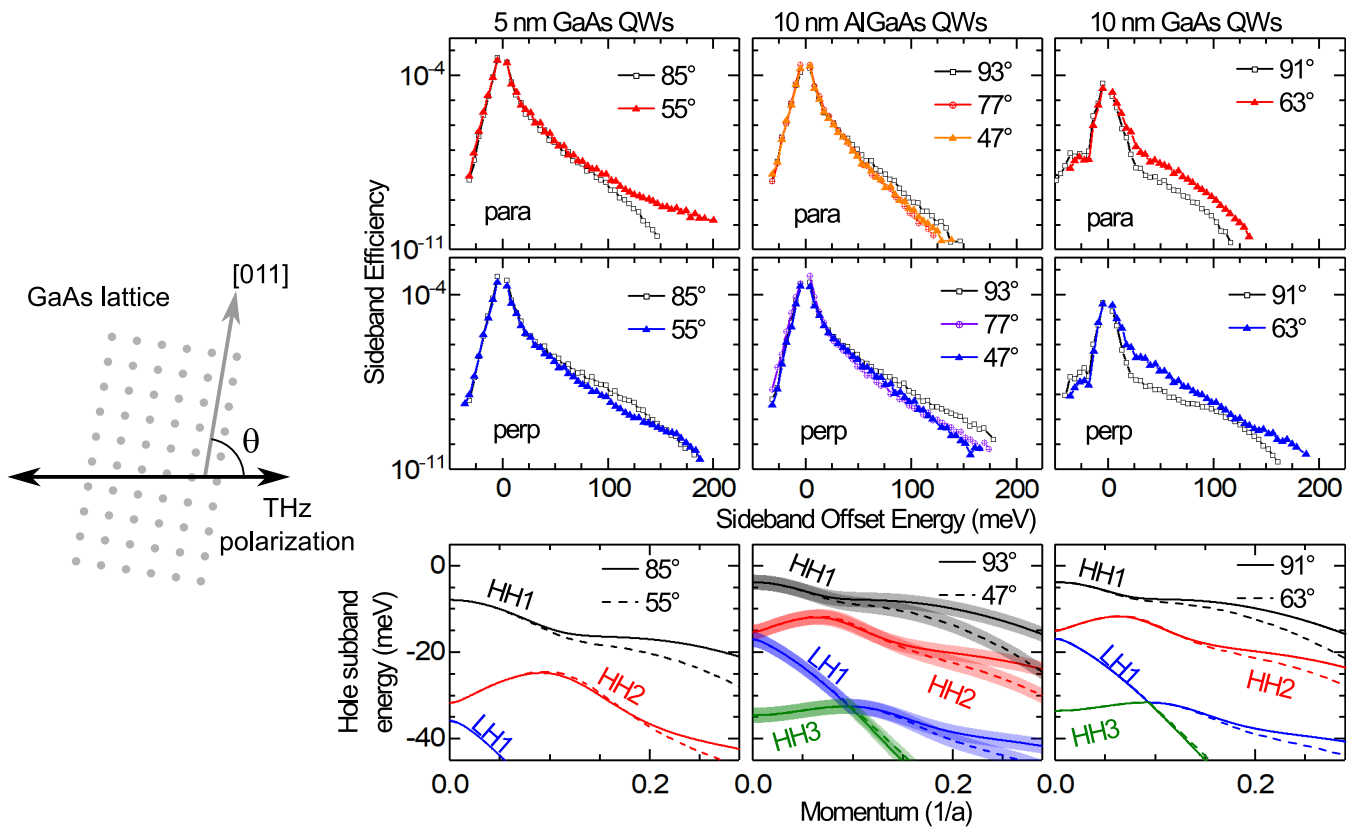}
		\caption{HSG spectra from all three samples for different lattice orientations, and calculated valence subband dispersion relations. The THz polarization was kept horizontal.  The angle $\theta$ between the [011] direction and the THz field is shown on the left side, with the propagation direction into the page. Experimental data shows sideband offset energy versus sideband (conversion) efficiency.  The empty black squares are replotted from Fig.~\ref{pol_comp}.  The valence subband dispersion relations are plotted along the directions of the THz electric field to elucidate the relation between the hole subbands and the sideband spectra.
		\textbf{5 nm GaAs QWs}: The relatively large avoided crossing between HH1 and HH2 is correlated with a relatively large energy required for the onset of dynamical birefringence (see Fig.~\ref{pol_comp}).  The dependence of HSG spectra on lattice orientation is relatively weak, except for the parallel-excited spectrum above 100 meV sideband offset energy.
		\textbf{10 nm AlGaAs QWs}: The avoided crossing between HH1 and HH2 is much smaller than for the 5 nm GaAs QW.  Broadening caused by alloy disorder is represented as a shaded strip on each curve in the dispersion relation.  The sideband conversion efficiencies show little dependence on lattice orientations.
		\textbf{10 nm GaAs QWs}: The hole dispersion relations are nearly identical to those for the 10 nm AlGaAs QW, but broadening from alloy disorder is absent in this sample, which has smaller disorder than the other two samples (see Fig.~\ref{appFigAbsorption} and the related discussions). The sideband spectra depend substantially on angle $\theta$. }
		\label{orientation}
	\end{figure*}

  Rotating the samples with respect to the THz polarization reveals that HSG is sensitive to the band structure along the direction of electron and hole motion in the 5 and 10 nm GaAs QWs, see Fig.~\ref{orientation}~\footnote{The crystal orientations of the samples were confirmed within the 4-fold symmetry by x-ray diffraction (XRD) experiments.}. On the left side of Fig.~\ref{orientation}, the orientation of the crystal lattice is defined by the angle $\theta$ between the [011] axis of the GaAs lattice and the THz electric field. The remaining panels of Fig.~\ref{orientation} plot sideband conversion efficiencies for parallel- and perpendicular-excited sidebands from all three samples with different values of $\theta$. The lowest conduction subband is approximately parabolic around the band minimum, so instead we focus on the valence subbands. The dispersion relations for the three highest valence subbands are plotted below the data on sideband conversion efficiency for each sample and value of $\theta$ measured (see Supplementary Material~\footnote{\label{band_model} See ``Band Model'' in Supplementary Materials for details of band structure calculation.} for the band structure calculation). While comparing HSG spectra with features in the hole dispersion relations, it is important to note that holes carry only 10-30\% of the total kinetic energy of an electron-hole pair at a given quasi-momentum~\footnote{We estimate the fraction of the kinetic energy in the hole by assuming that the dispersion relation of the hole band lie between the parabolic dispersion relations associated with heavy hole ($m_{hh}=0.51 m_0$) and light hole ($m_{lh}=0.082 m_0$), and the electron band is parabolic with effective mass $m_{e}=0.067 m_0$, where $m_0$ is the free electron mass. A few lines of algebra will show that, the fraction of the kinetic energy in the hole at recollision lies between $m_e/(m_e+m_{hh})\approx0.12$ and $m_e/(m_e+m_{lh})\approx0.45$. From the band calculation in Supplementary Material~\cite{Note4}, this range is about 10\% to 30\%. The total kinetic energy associated with a particular sideband is 5-10 meV smaller than the sideband offset energy because the NIR laser is tuned slightly below the band gap in order to resonantly drive excitons.}.
	
  The 5 nm GaAs sample has the strongest quantum confinement, which is correlated with having the largest energy required for the onset of dynamical birefringence. The spacing between the two highest heavy hole (HH) subbands, HH1 and HH2, is approximately 25 meV at zero momentum, as shown in the bottom left of Fig.~\ref{orientation}.  The subbands along the 55$^\circ$ and 85$^\circ$ directions are nearly indistinguishable in HH1 and HH2 subbands until a $\sim\!10$ meV wide avoided crossing, centered at a hole kinetic energy of 12 meV, near a quasi-momentum of about 0.1, in units of $1/a$ where $a$ is the lattice constant of bulk GaAs. Above 0.1 $1/a$, the gap between the HH1 and HH2 subbands is larger for the 85$^\circ$ than for the 55$^\circ$ orientation. The sideband conversion efficiencies for the parallel-excited sidebands are nearly indistinguishable until a sideband offset energy of about 100 meV. Above this sideband offset energy, the sideband conversion efficiency is larger for the orientation with the smaller avoided crossing (55$^\circ$). The sideband conversion efficiencies for the perpendicular-excited sidebands show much weaker dependence on sample orientations.

  The 10 nm GaAs sample has much weaker quantum confinement compared to the 5 nm GaAs sample, with a $\sim\!10$ meV spacing between HH1 and HH2 subbands at zero momentum, as shown in the bottom right of Fig.~\ref{orientation}. As in the 5 nm GaAs sample, the HH1 and HH2 subbands along the two directions here are nearly indistinguishable until an avoided crossing with a $\sim\!5$ meV gap, centered at hole kinetic energy of about 5 meV, near momentum 0.08 $1/a$. Above 0.08 $1/a$, the gap between HH1 and HH2 subbands is slightly larger for the 91$^\circ$ than for the 63$^\circ$ orientation. The sideband conversion efficiencies for both the parallel- and perpendicular-excited sidebands show a very strong dependence on lattice orientation above sideband offset energy of about 20 meV, and are larger for the 63$^\circ$ orientation.

  The 10 nm AlGaAs sample has nearly identical subband spacings as the 10 nm GaAs sample, but much stronger quenched disorder because of alloy scattering that is not present in either GaAs sample. The influence of disorder is represented as broadenings of the subbands in the lower center part of Fig.~\ref{orientation}. Interestingly, while the 10 nm AlGaAs and 10 nm GaAs samples showed birefringence above roughly the same offset energy, the sideband conversion efficiencies in the 10 nm AlGaAs sample depend very little on the orientations of the lattice. The persistence of dynamical birefringence in the face of alloy disorder is striking, and suggests that dynamical birefringence is related to local electronic structure, and not simply to crystal symmetry. We speculate that, at the length scale of a recollision (a few tens of nm), local avoided crossings persist in the presence of alloy disorder, so the THz-induced birefringence is similar in the 10 nm AlGaAs and GaAs samples. At the length scale of the 200 $\mu$m NIR spot, however, the ensemble of recollisions samples many different local band gaps. In the ensemble, the four-fold symmetry of the band structure may be averaged out to nearly cylindrical symmetry without masking the structure of the Bloch wave functions.

  The experimental results of this section are consistent with the notion that the direction of the THz electric field defines the birefringent axis, and that the effects of this \textit{dynamical} birefringence are related to the band structure. The 5 nm GaAs sample, with the narrower, more strongly confined wells, shows the weakest dynamical birefringence: the energy required for the onset of dynamical birefringence is the highest, and the contrast between the intensities of parallel- and perpendicular-excited sidebands is the weakest. The two 10 nm samples, despite having very different dependences of HSG spectra on crystal orientations, have similar degrees of dynamical birefringence. Before investigating the experimental results further, we consider possible physical mechanisms for this dynamical birefringence and develop a theoretical model to describe the phenomenon.
		
\section{Theoretical analysis} \label{TheoryModels}
  \subsection{Berry physics in HSG} \label{berry_physics}
  In a three-step model of high-order sideband generation (HSG), the first step is the creation of electron-hole pairs whose initial state is determined by the polarization of the excitation laser; the second step is the acceleration of electrons and holes along $k$-space trajectories in the conduction and valence bands; and the third step is the emission of a photon whose polarization state is determined by the final state of the electron-hole pair. Therefore, the polarization dependence of HSG is closely related to the variation of the internal states (including spin and orbital states) in the Brillouin zone, or the Berry connection. The Berry connection is defined as $\vec{R}_{mn}=\langle u_{m,\mathbf{k}}|i\partial_{\mathbf{k}}|u_{n,\mathbf{k}}\rangle$, where $|u_{n,\mathbf{k}}\rangle$ is the cellular function (internal state) of the $n$th band or subband at quasi-momentum ${\mathbf k}$. The Berry connection is in general non-Abelian and is Abelian if $\vec{R}_{mn}=\mathbf{0}$ for all $m\ne n$. As an excitation moves in the reciprocal space, Berry connections are accumulated as a Berry phase. Being gauge-dependent, the Berry connection is not a physical quantity. A gauge-independent quantity that characterizes the variation of cellular functions is the Berry curvature, which is defined as $\mathbb{F}^{\gamma} = \frac{1}{2} \varepsilon^{\alpha \beta \gamma} \mathbb{F}^{\alpha \beta}$ with $\mathbb{F}^{\alpha \beta}=i\left[D_{k_{\alpha}},D_{k_{\beta}}\right]$, where $D_{\mathbf{k}}=\partial_{\mathbf{k}}-i\vec{R}$ ($\alpha,\beta,\gamma=x,y,z$). In the Abelian case, the relation between the Berry connection and Berry curvature reduces to $\vec{\mathbb{F}}=\partial_{\mathbf{k}}\times\vec{R}$, similar to that between the vector potential and magnetic field in electromagnetism.
  
  In a band insulator with both time-reversal and inversion symmetries, and in-plane dipole matrix elements being cylindrically symmetric at a quasi-momentum $\mathbf{k}$ and nonzero only between valence and conduction bands, there should be no dynamical birefringence in HSG if the Berry connection is zero. It is convenient to describe the radiation on the basis $\sigma^{\pm}$ ($\sigma^{\pm}$ correspond to photons with angular momentum $\pm 1$), and we will call the corresponding components $\sigma^{\pm}$ photons. If the Berry connection is zero, the cellular functions for each energy band and the dipole matrix elements will be the same for all Bloch wave vectors. Thus, assuming zero intraband, inter-valence-band and inter-conduction-band dipole matrix elements, electron-hole pairs associated with different cellular functions will be completely decoupled. In this case, in the acceleration process, an electron-hole pair can only accumulate a dynamic phase and a dephasing factor, but no Berry phase. Without Berry phases, the recombination of an electron-hole pair created by a $\sigma^{+}$ NIR photon can only produce a $\sigma^{+}$ sideband photon, which carries the dynamic phase and the dephasing factor of the electron-hole pair. It is similar for a $\sigma^{-}$ NIR photon. Due to time-reversal and inversion symmetries, for each electron-hole pair created by a $\sigma^{+}$ NIR photon, there is always another one that can be created by a $\sigma^{-}$ NIR photon with the same dynamic phase and dephasing factor. Therefore, in the band insulator described above, a zero Berry connection implies that the amplitudes of the sidebands are proportional to the exciting NIR laser, which means rotating a linearly polarized NIR laser has no effect on the sideband intensity. For the mathematical details and generalizations, see Appendix~\ref{zero_berry}. The (100) GaAs QWs are band insulators that satisfy the conditions outlined above \footnote{Firstly, the QWs have both time-reversal and inversion symmetries. Secondly, when the QWs are resonantly excited, the electron-hole pairs that dominate the physics are created around $\mathbf{k}=\mathbf{0}$, where the in-plane transition dipole moments are of cylindrically symmetric forms $\sigma^{\pm}\equiv\pm({\hat{X} \pm i\hat{Y}})/{\sqrt{2}}$ ($\hat{X}$ and $\hat{Y}$ are unit vectors along $[010]$ and $[001]$, respectively). The last criterion, that there are no intraband, inter-valence-band or inter-conduction-band dipole transitions, is satisfied because the relevant cellular functions of GaAs have definite parities, which are even for the conduction bands and odd for the valence bands.}, so a nonzero Berry connection is essential to explain the observation of dynamical birefringence. 
		
  Dynamical birefringence arises as a result of quantum interference between electron-hole recollision pathways associated with different Berry phases in systems like the (100) GaAs QWs because of nonzero Berry connection. In the THz field, an electron-hole pair excited by a $\sigma^{+}$ NIR photon can evolve into a state whose in-plane transition dipole moment is a linear combination of $\sigma^{+}$ and $\sigma^{-}$. Thus, a sideband generated from a $\sigma^{+}$ NIR photon is of the form $a\sigma^{+}+b\sigma^{-}$, where the energy levels, the dephasing rates and the Berry curvatures are coded in the coefficients $a$ and $b$, which depend on the direction of the electron and hole motion. Similarly, for the same order of sideband generated from a $\sigma^{-}$ NIR photon, the radiation has the form $a'\sigma^{-}+b'\sigma^{+}$. With nonzero Berry connection, the polarization states of the sidebands can be quite different from that of the NIR laser. For an NIR laser linearly polarized along $e^{-i2\Psi}\sigma^{+}-\sigma^{-}$, each sideband is of the form $(e^{-i2\Psi}a-b')\sigma^{+}+(e^{-i2\Psi}b-a')\sigma^{-}$, a sum of interfering quantum amplitudes. The norm of that sum---the electric field amplitude of a particular sideband---depends on the polarization angle $\Psi$, as do the experimentally-observed sidebands. This analysis also suggests that sidebands should depend on the lattice orientation with respect to the THz field as observed, and, in general, should be elliptically polarized.
				
  \subsection{Semiclassical picture}\label{semiclassical picture}
  To construct a physical picture of dynamical birefringence, we establish a semiclassical theory with a non-Abelian Berry connection using the saddle-point method, which has been successfully used to construct semiclassical theories for both HHG~\cite{Lewenstein1994} and HSG~\cite{Yang2013,Yang2014a,Yang2014,yang2015geometric}.
	    
  We first model the band structure of the GaAs QWs by using the envelope function approximation based on the six-band Kane Hamiltonian~\cite{Kane1957}. Thus the basis we use to describe the electronic states contains combinations of envelope functions and the following cellular functions
  \begin{align}
	\Ket{u_1} &= \Ket{S, \uparrow}, \quad \Ket{u_2} = \Ket{S, \downarrow},\\
	\Ket{u_3} &= -\frac{1}{\sqrt{2}} \Ket{(X+iY), \uparrow} = \Ket{\frac{3}{2}, +\frac{3}{2}},\\
	\Ket{u_4} &= -\frac{1}{\sqrt{6}} \left[\Ket{(X+iY), \downarrow} - 2\Ket{Z, \uparrow}\right] = \Ket{\frac{3}{2}, +\frac{1}{2}},\\
	\Ket{u_5} &= \frac{1}{\sqrt{6}} \left[\Ket{(X-iY), \uparrow} + 2\Ket{Z, \downarrow}\right] = \Ket{\frac{3}{2}, -\frac{1}{2}},\\
	\Ket{u_6} &= \frac{1}{\sqrt{2}} \Ket{(X-iY), \downarrow} = \Ket{\frac{3}{2}, -\frac{3}{2}},
  \end{align}
  where $\Ket{S}$ belongs to the irreducible representation $\Gamma_1$ of $T_d$ symmetry group, $\Ket{X},\Ket{Y},\Ket{Z}$ belong to $\Gamma_4$, and $\Ket{\uparrow}, \Ket{\downarrow}$ are eigenvectors of Pauli matrix $\sigma_z$ in spin space. The eigenstates $\Ket{u_1},\Ket{u_2}$ are usually called the electron components. The eigenstates $\Ket{u_3},\Ket{u_4},\Ket{u_5},\Ket{u_6}$ can also be labeled by the quantum numbers of spin-$\frac{3}{2}$. The eigenstates $\Ket{u_3}$, $\Ket{u_6}$ are usually called the heavy-hole components, while $\Ket{u_4}$, $\Ket{u_5}$ are the so-called light-hole components.  In bulk GaAs, the energy gap is about $1.55$ eV, so we neglect the coupling between the electron and hole components and apply the effective-mass approximation for the conduction bands. Under the hard-wall approximation, the envelope functions are sinusoidal:
  \begin{equation}
	f_n(z)=\sqrt{\frac{2}{L}} 
	\begin{cases}
		\cos\left(\frac{2m-1}{L}\pi z \right)  &: n = 2m-1 \\
		\sin\left(\frac{2m}{L}\pi z \right)  &: n = 2m
	\end{cases},
  \end{equation}
  where $m=1,2,...$, $L$ is the well width, and $f_n(z)$ is odd/even as a function of $z$ when $n$ is even/odd. We denote the basis as $f_n \Ket{S, \uparrow}$, $f_n\Ket{S, \downarrow}$ for the electron components, and $f_n\Ket{\frac{3}{2}, \pm\frac{3}{2}}$, $f_n\Ket{\frac{3}{2}, \pm\frac{1}{2}}$ for the hole components. In our model, zero Berry connection is assumed for the conduction subbands, while heavy hole-light hole coupling induces a non-Abelian Berry connection in the valence subbands. See Supplementary Material~\cite{Note4} for more details about band structure calculations.
	    
  To study the simplest model that is expected to capture the main physics of HSG in the GaAs QWs, we will consider only the lowest conduction subband and the highest two valence subbands. As the QWs have both time-reversal and inversion symmetries, we choose the eigenstates in each subband as pairs related by a time-reversal transformation and an inversion. The cellular functions of the lowest conduction subband (E1 subband) are denoted as $\Ket{\text{E}_{1,\uparrow}}\equiv f_1 \Ket{S, \uparrow}$ and $\Ket{\text{E}_{1,\downarrow}}\equiv f_1 \Ket{S, \downarrow}$, which are $k$-independent under the assumption of zero Berry connection. For the $j$th highest valence subband (HH$j$ subband, $j=1,2$), the cellular functions are denoted as $\Ket{\text{HH}_{j,\uparrow}}_{\mathbf{k}}$ and $\Ket{\text{HH}_{j,\downarrow}}_{\mathbf{k}}$. With non-Abelian Berry connection, the states $\Ket{\text{HH}_{1,\uparrow}}_{\mathbf{k}}$ and $\Ket{\text{HH}_{2,\downarrow}}_{\mathbf{k}}$ are linear combinations of $f_{2m-1}\Ket{\frac{3}{2}, +\frac{3}{2}}$, $f_{2m}\Ket{\frac{3}{2}, -\frac{3}{2}}$, $f_{2m-1}\Ket{\frac{3}{2}, -\frac{1}{2}}$, $f_{2m}\Ket{\frac{3}{2}, +\frac{1}{2}}$ ($m=1,2, ...$), with $k$-dependent coefficients. Thus there are only four types of electron-hole pairs involved that have nonzero transition dipole moments: $f_1 \Ket{S, \uparrow}$-$f_1\Ket{\frac{3}{2}, +\frac{3}{2}}$, $f_1 \Ket{S, \uparrow}$-$f_1\Ket{\frac{3}{2}, -\frac{1}{2}}$, $f_1\Ket{S, \downarrow}$-$f_1\Ket{\frac{3}{2}, -\frac{3}{2}}$ and $f_1 \Ket{S, \downarrow}$-$f_1\Ket{\frac{3}{2}, +\frac{1}{2}}$ (see Supplementary Material~\footnote{\label{dyna_equation}See ``Dynamical Equation'' in Supplementary Materials for a discussion on the dynamics of electron-hole pairs}). The electron-hole pairs involving a $f_1\Ket{S, \uparrow}$ component are decoupled with those containing $f_1\Ket{S, \downarrow}$, and we can divide the electron-hole pairs into two groups, which are related to each other by a time-reversal transformation and an inversion. For one group, we have electron-hole pairs $\Ket{\text{E}_{1,\uparrow}}$-$\Ket{\text{HH}_{1,\uparrow}}_{\mathbf{k}}$, $\Ket{\text{E}_{1,\uparrow}}$-$\Ket{\text{HH}_{2,\downarrow}}_{\mathbf{k}}$, and for the other group, we have $\Ket{\text{E}_{1,\downarrow}}$-$\Ket{\text{HH}_{1,\downarrow}}_{\mathbf{k}}$, $\Ket{\text{E}_{1,\downarrow}}$-$\Ket{\text{HH}_{2,\uparrow}}_{\mathbf{k}}$.
	    
  With the band model established, we start the saddle-point analysis by writing the amplitude of the $n$th order sideband in the following standard path integral form 		
  \begin{align}
	\mathbf{P}_n &= \sum_{s=\uparrow,\downarrow} \int_{-\infty}^{+\infty}dt \int^t_{-\infty}dt' \int\frac{d\mathbf{P}}{(2\pi)^2} \int D[\phi^{\dag}_s,\phi_s]  \notag \\
	&\quad {\mathbf{D}}_s^{\dag}[\mathbf{k}(t)] \phi_s[\mathbf{k}(t)]e^{\frac{i}{\hbar}\mathbb{S}_s}\phi^{\dag}_s[\mathbf{k}(t')]{\mathbf{D}}_s[\mathbf{k}(t')] \cdot \mathbf{F}_\text{NIR}, \label{eqnSemiclassical}
  \end{align}  
  where $s=\uparrow,\downarrow$ labels the two groups of electron-hole pairs, $\hbar\mathbf{k}(t) = \hbar\mathbf{P} - e\mathbf{A}(t)$ is the kinetic momentum with $\hbar\mathbf{P}$ being the canonical momentum and $-\dot{\mathbf{A}} = \mathbf{E}_\text{THz}(t) = \mathbf{F}_\text{THz} \cos(\omega t)$ the THz electric field with frequency $\omega$, $\mathbf{E}_\text{NIR}(t) = \mathbf{F}_\text{NIR} e^{-i\Omega t}$ is the electric field of the NIR laser under the rotating wave approximation with frequency $\Omega$, $\mathbf{D}_{\uparrow}[\mathbf{k}(t)]$ and $\phi_{\uparrow}$ are respectively a two-component dipole vector and an $\text{SU}(2)$ functional field corresponding to electron-hole pairs $\Ket{\text{E}_{1,\uparrow}}$-$\Ket{\text{HH}_{1,\uparrow}}_{\mathbf{k}}$ and $\Ket{\text{E}_{1,\uparrow}}$-$\Ket{\text{HH}_{2,\downarrow}}_{\mathbf{k}}$, $\mathbb{S}_{\uparrow} = \int_{t'}^{t}\mathbb{L}_{\uparrow}[\mathbf{k}(t''),\dot{\mathbf{k}}(t''),\phi_{\uparrow},\dot{\phi_{\uparrow}}]dt'' + \hbar \Omega (t-t') + n \hbar \omega t$, with $\mathbb{L}_{\uparrow} = i\hbar\phi^{\dag}_{\uparrow}\dot{\phi}_{\uparrow}-\phi^{\dag}_{\uparrow}[\Lambda(\mathbf{k})-e\mathbf{E}_\text{THz}(t)\cdot \vec{R}_{\uparrow}(\mathbf{k})]\phi_{\uparrow}$, $\Lambda(\mathbf{k})=diag\{E_{\text{cv},1}(\mathbf{k}), E_{\text{cv},2}(\mathbf{k})\}$ is a diagonal matrix with $E_{\text{cv},j}(\mathbf{k})$ being the energy level difference between E1 subband and HHj subband, $\vec{R}_{\uparrow}(\mathbf{k})$ is the non-Abelian Berry connection, and $D[\phi^{\dag}_{\uparrow},\phi_{\uparrow}]$ is the functional measure (similar for $s=\downarrow$) (see Supplementary Material~\cite{Note7} for the derivation). The dephasing rate is neglected to simplify the picture and will be discussed in later sections. Variations with respect to $\mathbf{k}$ and $\phi^{\dag}_s$ respectively give the following saddle-point equations
	\begin{align}
		0 &= \int_{t'}^t\{\phi^{\dag}_s[\mathbf{k}(t'')]\frac{1}{\hbar}\left[D^s_{\mathbf{k}},\Lambda[\mathbf{k}(t'')] \right]\phi_s[\mathbf{k}(t'')] \notag \\
		&\quad - \dot{\mathbf{k}}(t'') \times \phi^{\dag}_s[\mathbf{k}(t'')]\vec{\mathbb{F}}_s[\mathbf{k}(t'')]\phi_s[\mathbf{k}(t'')]\}dt'', \label{first_eqn}\\
		i\hbar\frac{d\phi_s}{dt} &= \Lambda[\mathbf{k}(t)] \phi_s - e\mathbf{E}_\text{THz}(t)\cdot \vec{R}_{s}[\mathbf{k}(t)]\phi_s, \label{second_eqn}
	\end{align}
	where $[D^s_{\mathbf{k}},\Lambda(\mathbf{k})]/\hbar$ is the covariant relative group velocity of the electron-hole pairs with $D^s_{\mathbf{k}}=\partial_{\mathbf{k}}-i\vec{R}_s$ the covariant derivative, and $\vec{\mathbb{F}}_s$ is the non-Abelian Berry curvature matrix defined as $\mathbb{F}^{\gamma}_s = \frac{1}{2} \varepsilon^{\alpha \beta \gamma} \mathbb{F}_s^{\alpha \beta}$ with $\mathbb{F}^{\alpha \beta}_s = i\left[D^s_{k_{\alpha}},D^s_{k_{\beta}}\right]$ ($\alpha,\beta,\gamma=x,y,z$). 
	
	Another two saddle-point equations concerning conservation of energy are obtained by variations with respect to $t$ and $t'$:
	\begin{align}
		\text{Re} \left[\frac{\mathbf{F}_\text{NIR} \cdot {\mathbf{D}}_s^{\dag}(t')\Lambda[\mathbf{k}(t')]\phi_s(t')}{\mathbf{F}_\text{NIR} \cdot {\mathbf{D}}_s^{\dag}(t')\phi_s(t')}\right] &= \hbar \Omega, \label{third_eqn}\\
		\text{Re} \left[ \frac{\hat{E}_{l} \cdot {\mathbf{D}}_s^{\dag}(t)\Lambda[\mathbf{k}(t)]\phi_s(t)}{\hat{E}_{l} \cdot {\mathbf{D}}_s^{\dag}(t)\phi_s(t)}\right] &= \hbar (\Omega + n\omega), \label{fourth_eqn}
	\end{align}
	which require that the weighted average energy of the electron-hole pair equals the photon energy of the NIR laser at the moment of excitation, and reaches the sideband photon energy at the moment of recollision. The weights are proportional to the transition dipole moments. For instance, if we write $\phi_{\uparrow}=(\phi_{\uparrow,1},\phi_{\uparrow,2})^T$, and define $\Ket{\phi_{\uparrow,1}}=\phi_{\uparrow,1}\Ket{\text{E}_{1,\uparrow}}\Ket{\text{HH}_{1,\uparrow}}_{\mathbf{k}}$, $\Ket{\phi_{\uparrow,2}}=\phi_{\uparrow,2}\Ket{\text{E}_{1,\uparrow}}\Ket{\text{HH}_{2,\downarrow}}_{\mathbf{k}}$, then the average energy can be cast in the form $\text{Re}[\frac{\braket{ \mu | \phi_{\uparrow,1}} E_{cv,1} + \braket{ \mu | \phi_{\uparrow,2}} E_{cv,2}}{\braket{ \mu | \phi_{\uparrow,1}} + \braket{ \mu | \phi_{\uparrow,2}}}]$, where $\bra{\mu} = \bra{g} er_0$ with $\Ket{g}$ being the ground state and $r_0$ the projection of the radius vector along a certain direction. 
	
	Armed with Eqs.~\ref{first_eqn}-\ref{fourth_eqn}, a semiclassical picture can be constructed following the three-step sequence of the sideband generation process. This picture is shown schematically in Fig.~\ref{theory_schematic}.  For a detailed look at calculated sideband trajectories, see Appendix~\ref{appSemiclassical}, and for the method of solving these equations, see Supplementary Material~\footnote{\label{semi_cal}See ``Semiclassical Calculation'' in Supplementary Materials for details on the semiclassical calculation.}.
	
	First, an incoming NIR photon resonant with the energy gap is decomposed into circularly polarized components, $\sigma^{\pm}\equiv\pm({\hat{X} \pm i\hat{Y}})/{\sqrt{2}}$ ($\hat{X}$ and $\hat{Y}$ are unit vectors along $[010]$ and $[001]$, respectively). The $\sigma^{-}$ component excites an $f_1\Ket{\frac{3}{2},+\frac{3}{2}}$ electron from the highest valence subband to the state $f_1\Ket{S,\uparrow}$ in the lowest conduction subband, creating a spin-up electron wave packet and a hole wave packet with angular momentum $-{3}/{2}$ in real space, while the $\sigma^{+}$ component creates the time-reversal counterpart electron-hole pair.
	\begin{figure*}
		\includegraphics{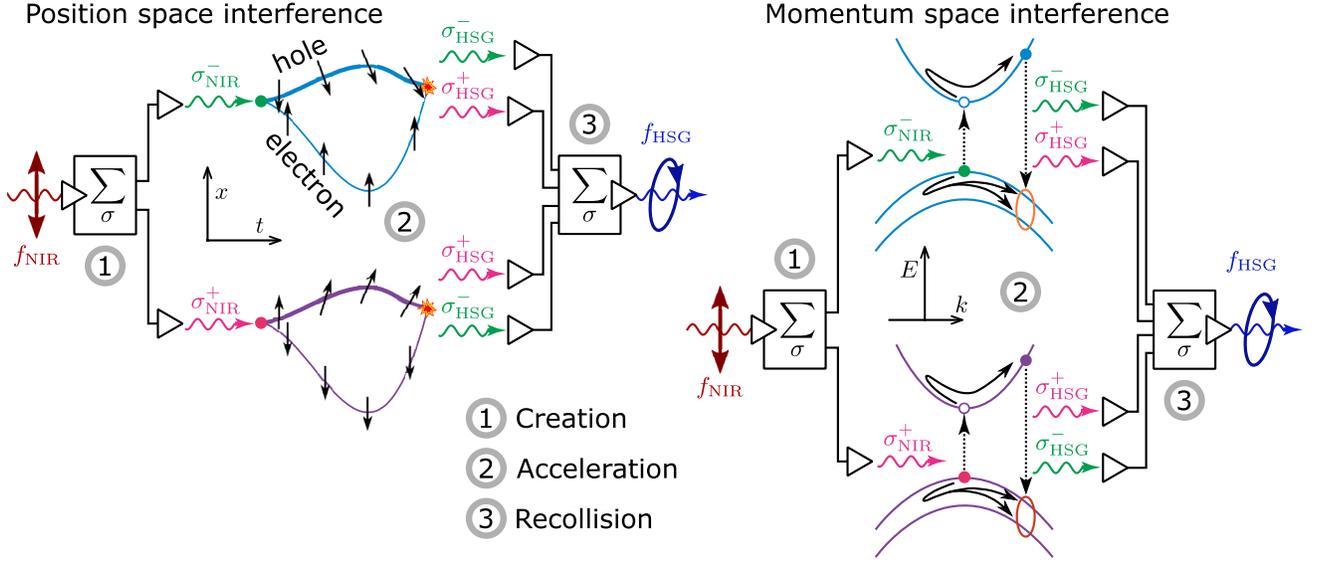}
		\caption{Schematic representation of 3-step model for high-order sideband generation in position space (left) and momentum space (right).  The figures represent a quantum interference process, with the upper (lower) arms associated with creation and dynamics of electron-hole pairs with total spin -1 (+1). 
		The linearly-polarized NIR excitation light is a superposition of a right and a left circularly-polarized component, $\sigma^\pm_\text{NIR}$. 
		In step (1), a quantum superposition of an electron-hole pair with total spin -1 (electron spin $+1/2$, heavy-hole spin $-3/2$) and an electron-hole pair with total spin angular momentum +1 (electron spin $-1/2$, heavy hole spin $+3/2$) is created. In step (2), the electrons and holes are accelerated by the THz field. In position space, the electron spinors (vertical arrows at 4 different times along ``electron'' trajectories, defined in Sect.~\ref{quantum_intereference}) do not rotate because Berry curvature is negligible, while the hole spinors rotate substantially under the influence of non-Abelian Berry curvature. In momentum space, the electron state remains confined to a single band during its acceleration, while the hole state mixes into nearby bands during its acceleration.  In step (3), electrons and holes recollide. Upon recollision, the hole state is a superposition of states (depicted by ovals in the right figure) in several different hole subbands with coefficients determined by the non-Abelian Berry curvature and the initial NIR excitation. Some of these states have allowed dipole transitions with the electron, and so create photons $\sigma^\pm_\text{HSG}$ that then interfere to generate the HSG emission at frequency $f_\text{HSG}$ with elliptical polarization. Rotating the NIR polarization changes the relative phase of the $\sigma^+_\text{NIR}$ and $\sigma^-_\text{NIR}$ photons and influences the output by changing how the emitted sideband photon states interfere, resulting in dynamical birefringence.}
		\label{theory_schematic}
	\end{figure*}
	
    In real space, as shown on the left of Fig.~\ref{theory_schematic}, after being created by the NIR laser, the electron and hole wave packets first move along opposite directions under the THz field, then are driven back and finally recollide with each other to generate sidebands. This recollision process is described by the first saddle point equation, Eq.~\ref{first_eqn}, in which the integrand can be written as the relative velocity of the electron and hole wave packets. With zero Berry connection, the velocity of the electron is the ordinary group velocity. For the hole, with non-Abelian Berry connection, the ordinary group velocity is replaced by a covariant group velocity, and there is a velocity component perpendicular to the THz field, which looks like the Lorentz force resulting from a $k$-space magnetic field (see Ref.~\cite{culcer2005coherent,sundaram1999wave}). Here, the Lorentz-like velocity of the hole is neglected for simplicity, because it is small for sidebands of order $n<60$ (see animations for $n=20,40,60$ in Supplementary Material~\footnote{\label{hole_evolution}See ``Hole Evolution'' in Supplementary Materials for animations of hole spinor evolution in real space for the $n=20,40,60$ sidebands.}). 

    In momentum space, the Berry physics shows up in a more direct way, as shown on the right side of Fig.~\ref{theory_schematic}. Driven by the THz field, each electron-hole pair moves around the Brillouin zone as $\hbar\dot{\mathbf{k}} = e\mathbf{E}_\text{THz}(t)$. The dynamics of the pseudo-spin $\phi_s$ is described by the second saddle-point equation, Eq.~\ref{second_eqn}. With zero Berry connection, the spin state of the electron remains the same. For the holes, we can consider $\phi_s$ as a pseudo-spin on the basis $\{\Ket{\text{HH}_{1,\uparrow}}_{\mathbf{k}}, \Ket{\text{HH}_{2,\downarrow}}_{\mathbf{k}}\}$, or $\{\Ket{\text{HH}_{1,\downarrow}}_{\mathbf{k}}, \Ket{\text{HH}_{2,\uparrow}}_{\mathbf{k}}\}$. During the acceleration process, a hole acquires a Berry phase, which is non-Abelian and induces Landau-Zener tunneling between the hole subbands. In the non-Abelian case, the dynamic phase and Berry phase are inseparable. In the language of spinors, the pseudo-spin $\phi_s$ is initially in the spin-up state, and precesses because of the non-Abelian Berry curvature.
	
	Finally, for the third step, consider the electron-hole pair created by a $\sigma^{-}$ NIR photon. At recollision, the electron is still in the state $f_1\Ket{S,\uparrow}$ while the hole has evolved into a superposition $\sum_{n,j}\eta_{n,j}f_n\Ket{u_j}$, where $\eta_{n,j}=\phi_{\uparrow,1}\alpha^{\uparrow,1}_{n,j}+\phi_{\uparrow,2}\alpha^{\uparrow,2}_{n,j}$ with $\alpha^{\uparrow,m}_{n,j}$ being the coefficient of $f_n\Ket{u_j}$ component in the cellular function of the $m$th subband. The non-Abelian geometric phase is carried by the pseudo-spin $(\phi_{\uparrow,1},\phi_{\uparrow,2})^T$. We have fixed the gauge at $\mathbf{k} = \mathbf{0}$ and the gauge is smoothed over the Brillouin zone, so $\alpha^{\uparrow,m}_{n,j}$ is uniquely determined. Except for the heavy-hole component $f_1\Ket{\frac{3}{2},+\frac{3}{2}}$ and the light-hole component $f_1\Ket{\frac{3}{2},-\frac{1}{2}}$, all other components in the hole wave packet cannot recombine with the electron state. Recombination of $f_1\Ket{S,\uparrow}$ electron with $f_1\Ket{\frac{3}{2},+\frac{3}{2}}$ hole and $f_1\Ket{\frac{3}{2},-\frac{1}{2}}$ hole respectively produces sideband components $\sigma^{-}$ and $\sigma^{+}$. Similarly, $\sigma^{+}$ photons can produce both $\sigma^{-}$ and $\sigma^{+}$ sideband photons. 

	Based on this picture, we explain the observed dynamical birefringence as a consequence of quantum interference between electron-hole recollision pathways injected with opposite spins. For an NIR laser linearly polarized along $\cos\Psi\hat{X} + \sin\Psi\hat{Y}$, both $\sigma^{+}$ and $\sigma^{-}$ components are present with a definite relative phase $\pi-2\Psi$. Consider the sideband component $\sigma^{+}$, which can be produced by both $\sigma^{+}$ and $\sigma^{-}$ NIR photons with different amplitudes denoted by $a$ and $b$. The sideband strength of the $\sigma^{+}$ component is proportional to $|e^{i(\pi-2\Psi)}a+ b|^2$. In our model, $a$ and $\sqrt{3}b$ are respectively the amplitudes of the heavy hole component $f_1\Ket{\frac{3}{2},-\frac{3}{2}}$ and light hole component $f_1\Ket{\frac{3}{2},+\frac{1}{2}}$ at recollision. We can immediately see that with a nonzero $b$, the sideband intensity should depend on the polarization of the NIR laser, because of the heavy hole-light hole coupling, or the nonzero Berry curvature. For simplicity, suppose the THz field is polarized along the [010] direction of the lattice and the NIR laser is linearly polarized parallel or perpendicular to the THz field. For the parallel case, the NIR laser is polarized along $\hat{X}\propto\sigma^{+}-\sigma^{-}$, so the sideband strength for the component $\sigma^{+}$ is  $I_{x,+}\propto|-a + b|^2$, while for the perpendicular case, since the NIR laser is polarized along $\hat{Y}\propto\sigma^{+}+\sigma^{-}$, the corresponding sideband strength is $I_{y,+}\propto|a + b|^2$. When the relative phase of $a$ and $b$ lies within $(-\frac{\pi}{2},\frac{\pi}{2})$, as in our cases, there will be $I_{y,+}>I_{x,+}$. For these special configurations, the sidebands are almost linearly polarized along the NIR laser, and the strength of each sideband is proportional to the one of the corresponding $\sigma^{+}$ component.
		
	Because of the Landau-Zener tunneling induced by the non-Abelian Berry curvature, the initially time-reversed electron-hole pairs can have nonzero total angular momentum at recollisions. This provides a mechanism for a linearly polarized NIR laser to produce elliptically polarized sidebands. In contrast, when both electrons and holes are restricted to a single subband, a linearly polarized NIR laser can only produce linearly-polarized sidebands, which may be rotated with respect to the polarization of the NIR laser by the Berry connection, which is Abelian in this case~\footnote{Rotations of linear polarization in HSG due to Abelian Berry phases for closed $k$-space paths have been discussed in Ref.~\cite{Yang2013}. The Berry phases in this paper are for open paths, which are not gauge-invariant of course, but the dipole matrix elements are also gauge-dependent and the whole polarization $\mathbf{P}_n$ is gauge-invariant.}. Note that Abelian Berry phases can induce elliptically polarized HSG from linearly polarized NIR laser, say, in the case when the electron-hole pair is created at more than one wave vector $\mathbf{k}$~\cite{yang2015geometric}. See Appendices~\ref{abelian_berry} and ~\ref{non_abelian_berry} for the mathematical details and generalizations.

	\subsection{Quantum simulation} \label{quantum_intereference}
    
    	To compare with experimental data, we then numerically simulate the evolution of interband polarizations, whose Fourier transforms give the HSG spectra. In the Heisenberg picture, when the Coulomb interaction is neglected, the dynamics of the electron-hole pairs is governed by
		\begin{align}
			i \hbar \frac{d \eta_{\mathbf{k}(t),s}}{dt} &= \Lambda[\mathbf{k}(t)]\eta_{\mathbf{k}(t),s} - e\mathbf{E}_\text{THz}(t) \cdot \vec{R}_{s}[\mathbf{k}(t)]\eta_{\mathbf{k}(t),s} \notag\\
			& - \mathbf{D}_{s}[\mathbf{k}(t)] \cdot \mathbf{E}_\text{NIR}(t) - i\hbar\gamma_{2}[\mathbf{k}(t)]\eta_{\mathbf{k}(t),s},\label{heisenberg}
		\end{align}
		where $\eta_{\mathbf{k}(t),\uparrow}=(c^{\dag}_{\text{H1},\uparrow,\mathbf{k}(t)}c_{\text{E},\uparrow,\mathbf{k}(t)}, c^{\dag}_{\text{H2},\downarrow,\mathbf{k}(t)}c_{\text{E},\uparrow,\mathbf{k}(t)})^{\text{T}}$, with $c^{\dag}_{\text{H1},\uparrow,\mathbf{k}(t)}$ and $c^{\dag}_{\text{H2},\downarrow,\mathbf{k}(t)}$ being creation operators corresponding to Bloch states $e^{i\mathbf{k} \cdot \mathbf{r}}\Ket{\text{HH}_{1,\uparrow}}_{\mathbf{k}}$ and $e^{i\mathbf{k} \cdot \mathbf{r}}\Ket{\text{HH}_{2,\downarrow}}_{\mathbf{k}}$, and $c_{\text{E},\uparrow,\mathbf{k}}$ an annihilation operator for an electron in the Bloch state $e^{i\mathbf{k} \cdot \mathbf{r}}\Ket{\text{E}_{1,\uparrow}}$ (similar for $\eta_{\mathbf{k}(t),\downarrow}$), $\gamma_{2}(\mathbf{k})=diag\{\gamma_{2,1}(\mathbf{k}),\gamma_{2,2}(\mathbf{k})\}$ is a diagonal matrix with each matrix element being a momentum-dependent dephasing rate due to phonon and impurity scattering. Low electron and hole densities are assumed, so that each electron-hole operator in $\eta_{\mathbf{k}(t),s}$ is approximately bosonic. Since the THz photon energy is much smaller than the energy gap and the NIR laser field is much weaker than the THz field, we neglect the THz field in the initial optical excitation process, while the NIR laser field is ignored when the electron-hole pairs are accelerated. After Eq.~\ref{heisenberg} is solved, the interband polarization is obtained as the expectation value of $\vec{\mathbb{P}}(t)=\sum_{s}\mathbf{D}_s^{\dag}[\mathbf{k}(t)]\eta_{\mathbf{k}(t),s}+H.c.$. In the numerical integration, we combine the leap-frog method with the Crank-Nicolson method, and only consider resonantly excited electron-hole pairs.

		In the next section, experiment and theory are compared.  In the hard-wall approximation used in the calculation, the height of the barrier is assumed to be infinite and the well-widths are enlarged.  In order to reproduce both the measured exciton peaks and the heavy hole-light hole exciton splitting, effective well-widths in the conduction and valence bands are assumed to be different. The effective well widths used in the calculation of the valence subbands (Fig.~\ref{orientation}) are 10.9 nm, 15.6 nm, 15.9 nm for the 5 nm GaAs QWs, the 10 nm AlGaAs QWs, and the 10 nm GaAs QWs respectively (see Supplementary Material~\cite{Note4} for more details).  Given the relatively severe approximations, and the small number of parameters adjusted, detailed  quantitative agreement between experiment and theory is not expected, but trends should be reproduced. See Supplementary Material~\cite{Note7} for the derivation of the dynamical equation and Supplementary Material~\footnote{See ``Quantum simulation'' in Supplementary Materials for details on the quantum simulation.} for the numerical method.

\section{Comparison of Experiment and Theory} \label{Comparison}

	\begin{figure}
		\includegraphics{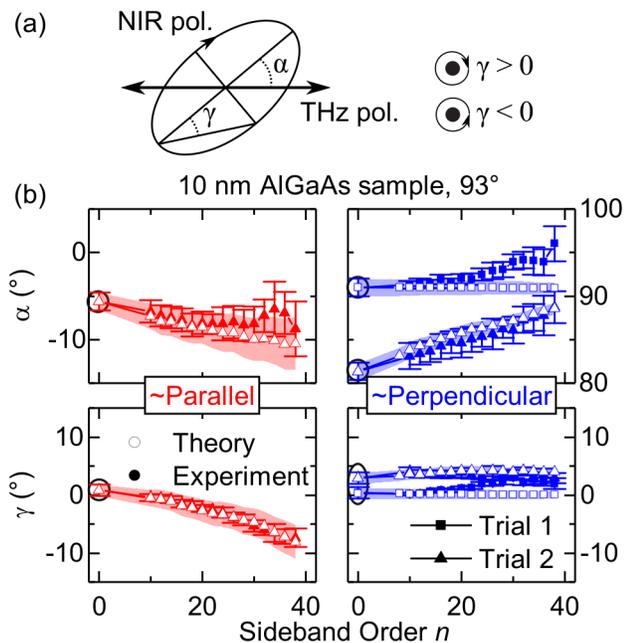}
		\caption{Polarization state of the sidebands in terms of the two ellipticity angles. (a) Definition of $\alpha$ and $\gamma$ for the polarization ellipse. (b) Experimental and theoretical values for the polarization angles. Experimental measurements are filled scatter points, theoretical calculations are empty scatter points. The shaded region is the error in the calculation propagated from error in the measured NIR laser polarization state. Polarization state measurements were performed on both 10 nm samples for both excitation geometries, and the nearly-perpendicular measurements were performed twice, labeled trials 1 and 2. For sideband polarimetry for the 10 nm GaAs sample, see Supplementary Material~\cite{Note2}. The angle of the [011] direction relative to the THz field is 93$^\circ$. The NIR laser polarization state is plotted as the order zero sideband and circled in black for each measurement. Overall, both theory and experiment agree that the polarization state of a given sideband is extremely sensitive to the NIR laser polarization state.}
		\label{pol_state}
	\end{figure}

	Both experiment and theory show that the polarization states of the sidebands are in general different from the polarization of the NIR laser, and change systematically with increasing sideband order. We measured the polarization states of the NIR laser and of the sidebands using a home-built Stokes polarimeter (see Supplementary Material~\cite{Note2} for experimental details). In all cases, we found the sidebands to be perfectly polarized, so that the polarization ellipse completely describes their polarization states. The polarization ellipse is parameterized by two angles: $\alpha$ is the angle the major axis makes with the dynamical optical axis defined by the direction of THz polarization, and $\gamma$ is the arctangent of the ratio of the semi-minor to semi-major axes, see Fig.~\ref{pol_state}(a). With the measured NIR laser polarization as an input (estimated error $\pm 1^\circ$), we also calculated the two ellipticity angles $\alpha$ and $\gamma$ associated with each sideband to compare with experiment.  
	
	For the 10 nm AlGaAs sample, with the NIR laser nearly linearly polarized parallel to the dynamical optical axis  ($\gamma\approx 0$, and $\alpha=-6^\circ$), both $\alpha$ and $\gamma$ of the sidebands rotate clockwise with increasing sideband order.  The calculated and measured $\alpha$ and $\gamma$ show the same trends and agree within the experimental error (See Fig.~\ref{pol_state}b, left).
	
	With the NIR laser within 10 degrees perpendicular to the dynamical optical axis, two measurement trials, together with associated calculations, show that the polarization states of the sidebands can depend sensitively on the polarization of the NIR laser (See Fig.~\ref{pol_state}b, right).  In trial 1, the NIR laser was nearly perfectly linearly polarized ($\gamma\approx 0$) at $\alpha$=91$^\circ$, which was oriented only 2$^\circ$ from the [011] direction. Theory predicts, in this special case, a nearly null effect--that the sidebands should all have polarizations that are extremely close to the polarization of the NIR laser.  This case is analogous to the case of linearly polarized light that is nearly parallel or perpendicular to the optical axis of a birefringent crystal, in which case the transmitted beam's polarization is nearly unchanged.  Indeed, measured values for $\gamma$ are close to calculated ones, showing nearly linear polarization at all measured orders ($\lvert\gamma\rvert<3^\circ$).  The measured $\alpha$ in trial 1 are within experimental error of the theoretical prediction up to about $n=20$, but are about 1 standard deviation above theory between 20 and 40. We note that, for the nearly perpendicular case, the error in calculated sideband polarization, which is propagated from the error in measured NIR laser polarization, decreases slightly with increasing order, making the comparison of experiment and theory in these cases more sensitive to approximations and systematic experimental errors than for the nearly parallel case (see Supplementary Material~\footnote{\label{j_matrix}See ``J matrix'' in Supplementary Materials for details on calculating the Jones matrix from experimental data.} for estimation of error propagation, and Supplementary Material~\cite{Note2} for a discussion of systematic errors).  
	
	In trial 2, the NIR laser was nearly linearly polarized ($\gamma$ =0) at an angle of $\alpha$=81$^\circ$ with respect to the dynamical optical axis.  Experiment and theory are in good agreement in this case: the sidebands in trial 2 remain nearly linearly polarized, and their polarization rotates counter-clockwise with increasing order, reaching nearly 90$^\circ$ at $n=40$. 
	
	Measurements and calculations were also performed for the 10 nm GaAs samples (see Supplementary Material~\cite{Note2}).  The sideband polarization states actually change more strongly with order than for the 10 nm AlGaAs sample.  However, perhaps because details of the band structure are more important in this cleaner sample, the quantitative agreement between experiment and theory is not as good as for the 10 nm AlGaAs sample.
	
\begin{figure}
		\includegraphics{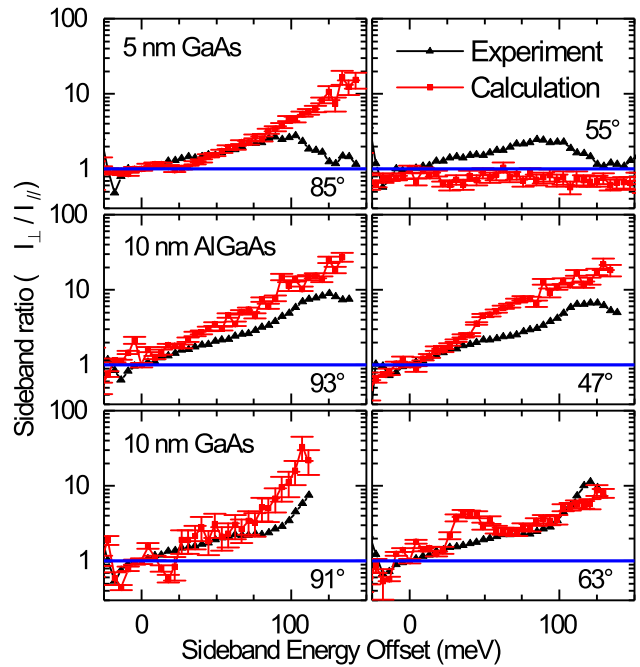}
		\caption{Comparison of dynamical birefringence in experiment and theory.  The ratio $I_\perp / I_\parallel$ is compared in all three quantum wells at different lattice angles. The nonzero Berry connection is responsible for deviations from unity. The quantum theory of sidebands only includes macroscopic dephasing, where the only $k$-dependence arises when the electrons (or holes) have enough energy to be elastically scattered into other bands (i.e. the dephasing rates are step functions). Comparing intensity ratios instead of absolute intensities largely cancels out the spin-independent scattering effect, which is assumed to be the dominant mechanism of scattering.  The blue line is the result if the Berry connection is assumed to be zero.}
		\label{expt_theory}
	\end{figure}

    In addition to accounting for dependence of sideband polarization on sideband order, the theory also accounts for the dependence of sideband intensity on NIR laser polarization. In order to largely factor out the effects of scattering on the intensities of high-order sidebands, we compare the theoretically-calculated and experimentally-measured ratios of the sideband intensities $I_{\bot}/I_{\parallel}$, where $I_{\bot}$ is for the case when the NIR laser field is perpendicular to the THz field and $I_{\parallel}$ for the case when the two fields are parallel, see Fig.~\ref{expt_theory}. 
    For the 5 nm GaAs sample at $85^{\circ}$ orientation, for sidebands of order $n \lesssim 46$ ($\approx100$ meV), the calculation and the experiment have an almost perfect match, increasing monotonically at the same rate. Above 100 meV, the experimentally measured ratio continues to increase, while the calculated one decreases. The trends for both theory and experiment for the 10 nm GaAs and AlGaAs QWs for all sample orientations are similar to those for the 5 nm GaAs QW at $85^{\circ}$, except that the experimentally-measured sideband ratios for the 10 nm GaAs QW, the cleanest sample, show some non-monotonic structure that is not present in the theory.  
    The experimental measurements for the 5 nm GaAs sample mounted at $55^{\circ}$ are quite different from all the others.  The measured ratios are all close to 1---there is, in this orientation, almost no dynamical birefringence!  The calculated sideband ratios for this orientation do not agree with the measured ones.  
	
	We do not understand the deviations between experiment and theory for the 5 nm GaAs QW mounted at $55^{\circ}$.  One possible cause is the hard-wall approximation.   The non-Abelian geometric phases depend on the band energy and the non-Abelian Berry connection due to heavy hole-light hole coupling, both determined by the well width (see Supplementary Material~\cite{Note4}).   As the effective well width decreases, the splitting of the valence subbands becomes larger, holes are more likely to remain in a single subband, and the Berry connection tends to zero. As discussed earlier, with a cylindrically symmetric dipole vector and zero Berry connection, the sideband polarization would be independent of the polarization of the NIR laser.  Therefore, the deviation of the ratio $I_{\bot}/I_{\parallel}$ for the 5 nm GaAs QWs at $55^{\circ}$ could be explained by an overestimation of Berry connection in the calculation (see Appendix~\ref{width_Berry} for more details).  However, a smaller Berry connection would also reduce the dynamical birefringence predicted at $85^{\circ}$, increasing deviations between experiment and theory along that direction.  A second possible cause is anisotropic scattering from fluctuations in well-width.  The 5 nm GaAs QW is actually more accurately described as consisting of rectangle-like “islands” that are 9, 10 or 11 monolayers thick, with the long direction parallel to the [$\bar{1}$10] direction~\cite{Neave1983}.   Scattering from these islands, which are likely smaller than the ~10 nm excitonic Bohr radius because the sample was grown without pauses at GaAs/AlGaAs interfaces~\cite{Tanaka1986},  may be different along the two directions measured.  Further investigations will be required to understand dynamical birefringence from such narrow QWs.

\section{Discussion} \label{Discussion}
\subsection{What about the ``plateau''?}\label{plateau}

	The classical three-step model predicts the existence of a ``plateau" in which the strengths of high-order sidebands~\cite{Liu2007} or harmonics~\cite{Lewenstein1994} depend relatively weakly on sideband order up to a cutoff.
	In atomic HHG, if the electron is launched on a valid recollision trajectory, there is little to stop it from recolliding with its parent ion, leading to high-order harmonics whose intensity varies little with order below the cutoff. In HSG, however, the electron and hole must interact with the lattice. Previously, scattering and dephasing were posited as the dominant mechanisms for the decrease in sideband strength with increasing sideband order~\cite{Banks2013,Langer2016}. The generalized three-step model presented here suggests that the Berry curvature should also contribute. When a hole that is initially in the HH1 state mixes into the nearby subbands upon acceleration, that hole is less likely to radiatively recombine with the electron upon recollision. In the GaAs QWs studied here, the probability of Landau-Zener tunneling between subbands increases dramatically with increasing sideband order (see Appendix~\ref{appSemiclassical}, Fig.~\ref{appFigSemiclassical}), and so the proportion of HSG-active holes decreases with increasing sideband order even in the absence of scattering. In general, even Abelian Berry curvature can cause electron-hole pairs to become a mixture of many components, most of which could be not HSG-active. We suggest that a clear plateau in HSG, one similar to the plateau observed in HHG from atoms, should only be expected in cases with nearly zero Berry curvature and weak scattering.

	\subsection{Why do disordered samples generate strong sidebands?}\label{disorder}
	
	When we began this study, we assumed that dephasing and decoherence were the dominant factors attenuating high-order sideband generation~\cite{Banks2013}, and we expected the highest sideband conversion efficiency to come from the least disordered sample. A careful look at Fig. 2 will show that the 10 nm GaAs sample, which was grown to have very smooth walls and has no Al atoms to cause band gap fluctuations, generates fewer and weaker sidebands than the more disordered 10 nm AlGaAs sample. We speculate that the hole Landau-Zener tunneling into dark states may explain this difference. In clean material, once the hole mixes into the HH2 subband, it is very likely to remain in that subband and be unable to radiatively recombine. In dirtier material, the disorder may suppress the coherent Landau-Zener tunneling and hence leave a larger component of the hole in the HH1 subband, from which it can radiatively recombine. Further theoretical work is necessary in order to fully understand the role of disorder in HSG.

	\subsection{A proposal for measuring band structure and Berry curvature}\label{proposal}
	
	We conclude the discussion section of this paper by explaining how the generalized three-step model presented here, in combination with experimental measurements like those presented here, can be used to determine band structures, Berry curvatures, and dephasing rates in a self-consistent way. This is significant because, although the dispersion relations of energy bands can be measured by Angle-Resolved Photoemission Spectroscopy (ARPES) and magnetotransport~\cite{Eisenstein1984}, and ARPES has been used to measure Berry phase in a special situation~\cite{hwang2011direct}, these techniques are not sensitive, to our knowledge, to Berry curvature. We note that it has been suggested that HHG may be used to measure Berry curvature~\cite{Liu2017}.
	
	To clarify the role of polarization in this proposed technique, we introduce a simple linear formalism that relates the electric field of the $n$th sideband to that of the NIR laser with a complex, $2\times2$ matrix we call a dynamical Jones matrix $\mathcal{J}_n$ ( $\mathcal{T}_n$ ) for a linearly (circularly) polarized basis, see Appendix~\ref{DynamicalJonesCalculus} for details. It is straightforward to measure this matrix experimentally. In fact, the ratio $I_{\bot}/I_{\parallel}$ plotted in Fig.~\ref{expt_theory} is simply $|{J_{yy,n}}/{J_{xx,n}}|^2$, although several more measurements are required to fully determine $\mathcal{J}_n$.  Theoretically, it is more natural to work in the circularly-polarized basis, and semiclassical or quantum theory can be used to calculate $\mathcal{T}_n$. The matrices $\mathcal{T}_n$ and $\mathcal{J}_n$ are related by a unitary transformation.  Good agreement between the experimental and theoretical values of this matrix then confirm a convincing understanding of the host material.
	
	There are likely many ways to solve the inverse problem of extracting band structures, Berry curvatures, and dephasing rates by comparing experimentally-measured and theoretically-calculated dynamical Jones matrices. One method is to establish a trial band model ~\cite{Vampa2015prl} and a trial dephasing model from which to calculate $\mathcal{T}_n$. The calculated $\mathcal{T}_n$ is then compared with the measured $\mathcal{T}_n$ and the band model is modified iteratively until the measured $\mathcal{T}_n$ is reproduced. We can start from low-order sidebands with a tight-binding model extended from the $\mathbf{k}\cdot\mathbf{p}$ theory as is used in this paper, in which the low energy physics is well described. For high-order sidebands, more high energy terms might be needed to better describe the large-$k$ behavior.

	The theoretical model that can reproduce the measured dynamical Jones matrix is not unique unless all components of the wave functions are optically active in the experiments. For example, in our experiments, the probability of the hole being in the HH2 subband is quite small near the zone center, where the Berry connection in this subband is hardly accumulated and is irrelevant.  In order to measure the Berry connection for HH2 subband near $\mathbf{k}=\mathbf{0}$, a stronger THz field or an NIR frequency resonant with that exciton should be used.

	Several experimental techniques can be used to improve the results of the self-consistent algorithm. For example, exciting with purely circularly polarized light will isolate any polarization changes to just one electron and hole species, measuring the non-Abelian Berry curvature more directly by eliminating the complex interference generated by linearly polarized light. Then, by tuning the THz frequency and field strength, the timescales for sidebands of the same order or offset energy can be changed to measure scattering rates with different time constants. As is done in Ref.~\cite{Langer2016}, one can also use NIR pulses instead of continuous waves. The advantage of using pulses is that it is possible to inject electron-hole pairs at particular phases of the THz field so as to initiate designated quantum trajectories.

	When the semiclassical theory works well or there is only a single quantum path for each spin sector, we can already measure the Berry phase in the Abelian case, where the dynamical Jones matrix can be approximated as
	\begin{equation}
			\mathcal{T}_n\propto
			e^{-\Gamma_d}e^{i\Gamma_D}
			\begin{pmatrix}
				e^{i\Gamma_B}\alpha_H & \frac{1}{\sqrt{3}}e^{-i\Gamma_B}\alpha^{*}_L\\
				\frac{1}{\sqrt{3}}e^{i\Gamma_B}\alpha_L & e^{-i\Gamma_B}\alpha^{*}_H
			\end{pmatrix},
		\end{equation}
	where $\Gamma_d$, $\Gamma_D$,and $\Gamma_B$ are respectively the dephasing factor, dynamic phase and the open-path Berry phase for the electron-hole pair created by an $\sigma^+$ NIR photon. $\alpha_H$ and $\alpha_L$ are the coefficients in the cellular function for the components $f_1|\frac{3}{2},-\frac{3}{2}\rangle$ and $f_1|\frac{3}{2},+\frac{1}{2}\rangle$. We can choose the gauge that $\alpha_H$ is real at $\mathbf{k}=\mathbf{0}$, which means through gauge smoothing, $\alpha_H$ can be made real all over the whole Brillouin zone. Since $\mathcal{T}_n$ can be determined from the techniques in this paper to within a phase factor, ${T_{--,n}}/{T_{++,n}}$ can be measured, which is just $e^{-2i\Gamma_B}$.
	
\section{Conclusion} \label{Conclusion}

	In conclusion, we have studied how the interplay between the relative orientations of the NIR laser polarization, THz polarization, and lattice affects HSG. We have measured HSG spectra of up to the 90th order and spanning over 200 meV, a bandwidth of over 12\% of the NIR laser frequency, by manipulating those relative orientations. We have shown conclusively that electrons and holes accelerate coherently through the lattice before recolliding. This coherence allows for interference between different electron and hole pathways, initialized by the NIR laser polarization and caused by non-Abelian Berry curvature in the hole subbands, and leads to large changes in sideband strength and sideband polarization state.

	In the next experiments, the observations discussed here should lead to a new generation of complete band structure measurement because HSG is inherently sensitive to both elements of the Bloch wavefunction, $e^{i\mathbf{k} \cdot \mathbf{x}} \ket{u_\mathbf{k}}$. By clever control of the NIR laser polarization, the THz frequency and field strength, and lattice orientation, a self-consistent algorithm can be developed for the direct measurement of the electron and hole dispersion relations, non-Abelian Berry curvatures, and even $k$- and $t$-dependent scattering rates in a broad class of materials.

\section{Acknowledgments}

	We would like to thank Garrett Cole, David Follman, and Paula Heu of Crystalline Mirror Solutions for teaching us the epitaxial transfer technique; John Leonard for teaching us ITO deposition techniques; Andrew Pierce for developing the methodology to measure the field-enhancement factor of the QW-sapphire-ITO system; and David Enyeart for maintaining, repairing, and assisting with the operation of the UCSB FEL. The UCSB FEL upgrade that made this work possible was funded by a Major Research Instrumentation (MRI) grant from the National Science Foundation, NSF-DMR 1126894. HB, DV and MSS, were funded by NSF-DMR 1405964. DV and MSS received additional support from the Office of Naval Research under grant N00014-13-1-0806. QW and RBL were supported by Hong Kong RGC and CUHK VC's One-Off Discretionary Fund.
	
	HBB and DCV conducted the experiments under MSS's supervision.  QW developed the theory and did all calculations under RBL's supervision.  SM grew the 10 nm AlGaAs and 5 nm GaAs samples under ACG's supervision.  LP grew the 10 nm GaAs sample.  MSS, QW, HBB, DCV, and RBL wrote the manuscript.

\appendix

\section{Sample absorption} \label{appAbsorption}

\begin{figure}
		\includegraphics{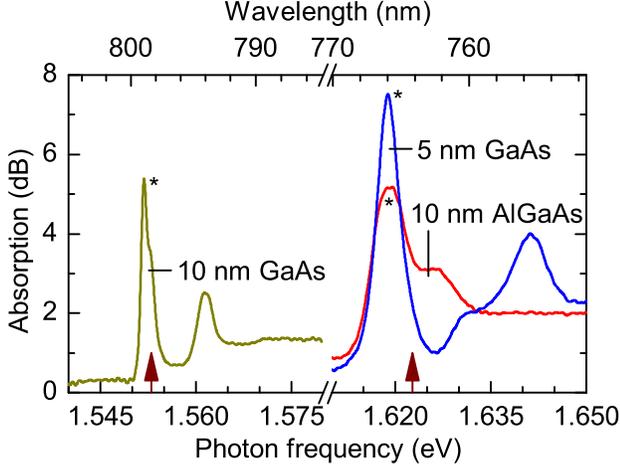}
		\caption{Optical absorption spectra of the three samples measured by differential transmission. In all three samples, the heavy hole exciton lines (starred peaks) and the light hole exciton lines (slightly weaker, blue-shifted by 10 or 25 meV depending on the well widths) are both clearly resolved. The measurements are performed at 15 K. In the 5 nm GaAs sample, the onset of the heavy hole 2D band gap is apparent at about 1.630 eV. The onset of the light hole 2D band gap is also evident (not pictured).  The smaller heavy hole-light hole splitting in the 10 nm samples masks the heavy hole 2D band gap in them, but, in the 10 nm GaAs sample, the onset of the light hole 2D band gap is apparent at 1.570 eV.  The NIR laser wavelength used for the HSG experiments are shown as dark red arrows.}
		\label{appFigAbsorption}
	\end{figure}

	Near-IR absorption measurements were performed on all samples using methods described in Supplementary Materials~\cite{Note2}. Several excitonic features are apparent in the absorption spectrum for each sample (see Fig.~\ref{appFigAbsorption}). The lower and higher energy peaks are assigned to the heavy-hole exciton (HHX) and light-hole exciton (LHX), respectively.  The splitting between the HHX and LHX peaks arises because quantum confinement breaks the HH-LH degeneracy at the top of the valence band.  In the 5 nm QWs, the HHX-LHX splitting is 25 meV. In both 10 nm QW samples, the HHX-LHX splitting is 10 meV, since larger well widths lead to smaller splitting due to weaker quantum confinement. The two 10 nm samples have different aluminum concentrations in the well region leading to the absorption differences. For the 10 nm AlGaAs sample, the 5\% Al content increases the 2D band gap so that the HHX absorption line is blue-shifted up to coincide with the HHX absorption line from the 5 nm GaAs sample. The 5 nm GaAs and 10 nm AlGaAs samples were produced by the same epitaxial growth as samples studied in Ref.~\cite{Banks2013}.
	
	The linewidths of the exciton lines probe the quenched disorder of the samples and so depend on the composition and growth conditions. The full width at half max (FWHM) of the HHX line in the 10 nm GaAs sample is 2.0 meV. In both the 5 nm GaAs sample and the 10 nm AlGaAs sample, it is 6.3 meV. There are two important sources of the inhomogeneous broadenings, well-width fluctuations and alloy disorder in the well region. Well-width fluctuations lead to a broadening that is inversely proportional to the well width, and so it is the dominant source of the inhomogeneous broadening for the 5 nm GaAs sample. If alloy disorder is modeled by Poisson-distributed aluminum content fluctuations over the 10-nm-diameter exciton wavefunction leading to local fluctuations in the band edge, we should expect an inhomogeneous excitonic linewidth broadening of about 5 meV for the 10 nm AlGaAs sample with 5\% aluminum concentration, consistent with the measured linewidth. The 10 nm GaAs sample, with wider and un-alloyed QWs, has the narrowest linewidth.

\section{Mathematics in Berry physics}

    To mathematically study the Berry physics in HSG, we investigate the amplitude of the $n$th order sideband in the form
	\begin{align}
		\mathbf{P}_{n} &= \frac{i}{\hbar} \int_{-\infty}^{+\infty}dt \int_{-\infty}^{t}dt' \int\frac{d \mathbf{P}}{(2\pi)^d} \notag \\ 
		&\quad e^{i(\Omega + n \omega) t} {\mathbf{D}}^{\dag} [\mathbf{k}(t)] e^{-\frac{i}{\hbar} \mathbb{S}}{\mathbf{D}}[\mathbf{k}(t')] \cdot \mathbf{E}_\text{NIR}(t'), \label{apphsg_amplitude}\\
		e^{-\frac{i}{\hbar} \mathbb{S}} &= \hat{T}\exp\{-\frac{i}{\hbar}\int_{t'}^t \mathbb{H}[\mathbf{k}(t'')]dt''\},
	\end{align}
	where $d$ is the dimension of the system, $\mathbb{H} (\mathbf{k}) = \Lambda(\mathbf{k})-e \mathbf{E}_\text{THz}(t) \cdot \vec{R}(\mathbf{k})+\mathbf{E}_\text{THz}(t)\cdot \mathbf{D}_{\text{int}}(\mathbf{k})- i\hbar\gamma_{2}(\mathbf{k})$, $\hat{T}$ denotes the time-ordering operator in the integration, and $\mathbf{D}_{\text{int}}(\mathbf{k})$ is a matrix describing the intraband, inter-valence-band and inter-conduction-band dipole matrix elements. We have assumed that the dephasing rates only depend on the band index and the quasi-momentum. The symbols used here are similar to those in Eq.~\ref{heisenberg} but in a more general sense that there could be more energy bands included. See Supplementary Material~\cite{Note7} for more details.

    \subsection{Zero Berry connection} \label{zero_berry}

        In a band insulator with zero intraband dipole matrix elements, if there are only one conduction band and one valence band involved in HSG, a zero Berry connection will imply that all sidebands should have the same polarization state, and the HSG spectra will be independent of the NIR laser polarization, except for an overall factor that is uniform for all orders of sidebands (the background optical birefringence of the sample). If the Berry connection is zero, then the dipole matrix elements are the same for all Bloch wave vectors. Since there are only one conduction band and one valence band involved, with zero intraband dipole matrix elements, we have $\mathbf{D}_{\text{int}}=\mathbf{0}$, so that $\mathbb{H} (\mathbf{k})$ is diagonal. In this case, the sideband amplitude (Eq.~\ref{apphsg_amplitude}) can be simplified to
    		\begin{align}
    			\mathbf{P}_n &= \frac{i}{\hbar}\int_{-\infty}^{+\infty}dt\int_{-\infty}^{t}dt'\int\frac{d\mathbf{P}}{(2\pi)^d} e^{i(\Omega + n\omega)t}\notag\\
    			&\quad  \sum_{j,g_j}e^{-\frac{i}{\hbar}\int_{t'}^t H_j[\mathbf{k}(t'')]dt''} \mathbf{d}^{*}_{g_j} \mathbf{d}_{g_j}\cdot\mathbf{E}_\text{NIR}(t'), \label{apphsg_amplitude_zeroberry}
    		\end{align}
    	summing over the electron-hole pairs from different bands and degenerate states (labeled by $j$ and $g_j$ respectively), with $H_j(\mathbf{k})=E_{\text{cv},j}(\mathbf{k})-i\hbar\gamma_{2,j}(\mathbf{k})$ and $\mathbf{d}_{g_j}$ being a constant dipole vector. If there are only one conduction band and one valence band, the label $j$ has only one value and the dynamic phases and dephasing factors associated with each $k$-space trajectory (fix $t$, $t'$ and canonical momentum $\mathbf{P}$ in Eq.~\ref{apphsg_amplitude_zeroberry}) are the same for all electron-hole pairs. Thus, we have $\mathbf{P}_n\propto \sum_{g_j}\mathbf{d}^{*}_{g_j} \mathbf{d}_{g_j}\cdot\mathbf{F}_\text{NIR}$, which means all sidebands have the same polarization state. In an HSG spectrum with logarithmic scales, a variation of the NIR laser polarization will only induce an overall change of sideband intensity.

        In a band insulator with more than two bands involved in HSG, even if the Berry connection is zero, sideband polarization states and degrees of dynamical birefringence can depend on the sideband order. In this case, electron-hole pairs can be created directly by the NIR laser from more than two bands, or can be first created by the NIR laser from two bands and then tunnel to other bands through inter-valence-band or inter-conduction-band transition dipole moments. For simplicity, we discuss the case when $\mathbf{D}_{\text{int}}=\mathbf{0}$. Suppose associated with each sideband, there are two electron-hole pairs from different energy bands with different dipole vectors (labeled by $j=1,2$) and a $k$-space trajectory. If the Berry connection is zero, we can still apply Eq.~\ref{apphsg_amplitude_zeroberry}, from which the amplitude of a sideband generated by the two electron-hole pairs has the form $\mathbf{P}_n\propto\sum_{j=1,2}\exp[i\Gamma_{D,j}-\Gamma_{d,j}]\mathbf{d}^{*}_j \mathbf{d}_j\cdot\mathbf{F}_\text{NIR}$, where $\Gamma_{D,j}=-\int_{t'}^t E_{cv,j}[\mathbf{k}(t'')]dt''/\hbar$ and $\Gamma_{d,j}=\int_{t'}^t \gamma_{2,j}[\mathbf{k}(t'')]dt''$ ($j=1,2$) are the dynamic phase and dephasing factor, which are in general not the same for different bands and depend on the sideband order. Therefore, the sideband polarization states and the degrees of dynamical birefringence should, in general, depend on the sideband order.

        In a band insulator with both time-reversal and inversion symmetries, and in-plane dipole matrix elements being cylindrically symmetric at a quasi-momentum $\mathbf{k}$ and nonzero only between valence and conduction bands ($\mathbf{D}_{\text{int}}=\mathbf{0}$), there should be no dynamical birefringence in HSG if the Berry connection is zero. As discussed above, Eq.~\ref{apphsg_amplitude_zeroberry} is valid in this case, and in addition, all in-plane transition dipole moments are cylindrically symmetric, i.e., $\mathbf{d}_{g_j}\propto\sigma^{+}$ or $\sigma^{-}$. Consider a recollision pathway along a $k$-space trajectory from $\mathbf{k}(t')$ to $\mathbf{k}(t)$ for an electron-hole pair with transition dipole moment $\mathbf{d}_{1}\propto\sigma^{+}$. The contribution of this pathway to the sideband amplitude is $C_0\exp[i\Omega(t-t') + in\omega t+i\Gamma_{D,1}-\Gamma_{d,1}]\sigma^{*}_{+} \sigma_{+}\cdot\mathbf{F}_\text{NIR}$, where $\Gamma_{D,1}$ and $\Gamma_{d,1}$ are the dynamic phase and dephasing factor respectively as defined in previous paragraph, and $C_0$ is a constant that does not depend on the choices of recollision pathways. Due to time-reversal and inversion symmetries, there is another recollision pathway for an electron-hole pair with the same $k$-space trajectory, band energy difference $E_{\text{cv},2}(\mathbf{k})=E_{\text{cv},1}(\mathbf{k})$ and dephasing rate $\gamma_{2,2}(\mathbf{k})=\gamma_{2,1}(\mathbf{k})$ but a complex conjugate dipole moment $\mathbf{d}_2=\mathbf{d}^{*}_1$. The contribution of this second pathway to the sideband amplitude is $C_0\exp[i\Omega(t-t') + in\omega t+i\Gamma_{D,1}-\Gamma_{d,1}]\sigma^{*}_{-} \sigma_{-}\cdot\mathbf{F}_\text{NIR}$. The sum of the contributions from these two recollision pathways is proportional to  $-(\sigma^{+}\sigma^{-}+\sigma^{-}\sigma^{+})\cdot \mathbf{F}_{\text{NIR}}=\mathbf{F}_{\text{NIR}}$. Thus, for such a band insulator, a zero Berry connection implies that the amplitudes of the sidebands are proportional to the exciting NIR laser, which means rotating a linearly polarized NIR laser has no effect on the sideband intensity, i.e., no dynamical birefringence. The proof above does not require the laser fields to be continuous waves (CWs).

        If the laser fields are CWs, then the statement in the previous paragraph is still valid for even order sidebands in the absence of inversion symmetry. Consider a recollision pathway along $k$-space trajectory $\mathbf{k}(t'')$ from $\mathbf{k}(t')=\mathbf{k}_0$ to $\mathbf{k}(t)=\mathbf{k}_e$ for an electron-hole pair with transition dipole moment $\mathbf{d}_{1}\propto\sigma^{+}$. The contribution of this pathway to the sideband amplitude is still $C_0\exp[i\Omega(t-t') + in\omega t+i\Gamma_{D,1}-\Gamma_{d,1}]\sigma^{*}_{+} \sigma_{+}\cdot\mathbf{F}_\text{NIR}$, with $C_0$, $\Gamma_{D,1}$ and $\Gamma_{d,1}$ defined the same as above. By a time-reversal transformation, we can find another recollision pathway for an electron-hole pair with transition dipole moment $\mathbf{d}_{2}=\mathbf{d}^{*}_{1}$ along $k$-space trajectory $\bar{\mathbf{k}}(\bar{t})=-\mathbf{k}(t'')$ from $-\mathbf{k}_0$ to $-\mathbf{k}_e$, where $\bar{t}=t''+\pi/\omega$, since a CW THz field changes its sign every half a period. The band energy difference and dephasing rate for this second electron-hole pair satisfy $E_{cv,2}(\mathbf{k})=E_{cv,1}(-\mathbf{k})$ and $\gamma_{cv,2}(\mathbf{k})=\gamma_{cv,1}(-\mathbf{k})$, so the dynamic phases and the dephasing factors associated with the two time-reversed pathways are the same. The contribution of this second pathway to the sideband amplitude is $(-1)^nC_0\exp[i\Omega(t-t') + in\omega t+i\Gamma_{D,1}-\Gamma_{d,1}]\sigma^{*}_{-} \sigma_{-}\cdot\mathbf{F}_\text{NIR}$. The sum of the contributions from these two recollision pathways is proportional to  $-[(-1)^n\sigma^{+}\sigma^{-}+\sigma^{-}\sigma^{+}]\cdot \mathbf{F}_{\text{NIR}}$. Therefore, even order sideband amplitudes are proportional to the NIR laser polarization, while, all odd order sidebands have the same polarization that is a mirror image of the NIR laser polarization, apart from the different intensities.

    \subsection{Abelian Berry connection}  \label{abelian_berry}

	In a band insulator with time-reversal and inversion symmetries, and in-plane dipole matrix elements being nonzero only between valence and conduction bands ($\mathbf{D}_{\text{int}}=\mathbf{0}$), when there are only two electron-hole recollision pathways (related by time-reversal and inversion symmetries with the same $k$-space trajectory) associated with each sideband, an Abelian Berry connection can only induce rotations of linear polarizations.
	If the Berry connection is Abelian, the matrices $\vec{R}(\mathbf{k})$, $\Lambda(\mathbf{k})$ and $\gamma_{2,j}(\mathbf{k})$ are all diagonal and commute with each other. Taking into account only two electron-hole recollision pathways related by time-reversal and inversion symmetries with a $k$-space trajectory from $\mathbf{k}(t')$ to $\mathbf{k}(t)$, we have from Eq.~\ref{apphsg_amplitude}
		\begin{align}
			\mathbf{P}_n\propto e^{i[\Omega(t-t')+n\omega t]}\sum_{j=1,2}e^{-\frac{i}{\hbar}\int_{t'}^t H_{j}[\mathbf{k}(t'')]dt''} \notag \\
			\times e^{i\Gamma_j[\mathbf{k}(t),\mathbf{k}(t')]} \mathbf{d}^{*}_j[\mathbf{k}(t)]\mathbf{d}_j[\mathbf{k}(t')]\cdot\mathbf{F}_\text{NIR}, \label{apphsg_amplitude_abelian}
		\end{align}
		where $\Gamma_j(\mathbf{k},\mathbf{k}') = \int_{\mathbf{k}'}^{\mathbf{k}}\vec{R}_{jj}(\mathbf{k}) \cdot d\mathbf{k}$ is the Berry phase for the electron-hole pair labeled by $j$, $\vec{R}_{jj}(\mathbf{k})$ is the corresponding Abelian Berry connection and $\mathbf{d}_j$ is a momentum-dependent dipole vector. As discussed in the case of zero Berry connection, the dynamic phases $\Gamma_D$ and dephasing factors $\Gamma_d$ are the same for these two pathways. With a suitable gauge choice, we make the dipole vectors for these two pathways be complex conjugates, and meanwhile the Berry phases be opposite~\cite{Yang2013}. Denote the dipole vector at $t_0$ for the first pathways as $\mathbf{d}_1[\mathbf{k}(t_0)]=a_{t_0}\sigma^{+}+b_{t_0}\sigma^{-}$, and the Berry phase it gains along the $k$-space trajectory as $\Gamma_B$. For an NIR laser linearly polarized along $\mathbf{F}_\text{NIR}=e^{-i\Psi}\sigma^{+}+e^{i\Psi}\sigma^{-}\propto \sin\Psi\hat{X}+\cos\Psi\hat{Y}$, the dipole coupling is $Q_0 \equiv \mathbf{d}_1 [\mathbf{k}(t')]\cdot \mathbf{F}_\text{NIR}=-(e^{i\Psi}a_{t'}+e^{-i\Psi}b_{t'})$ for the first pathway, and  $\mathbf{d}^{*}_1 [\mathbf{k}(t')]\cdot \mathbf{F}_\text{NIR}=-Q^{*}_0$ for the other. Thus, from Eq.~\ref{apphsg_amplitude_abelian}, we have $\mathbf{P}_n\propto e^{i\Gamma_B}Q_0(a_{t}\sigma^{+}+b_{t}\sigma^{-})-c.c.$, i.e., $\mathbf{P}_n\propto \rho(e^{-i\varphi}\sigma^{+} + e^{i\varphi}\sigma^{-})$, where $\rho e^{-i\varphi}=Q_0a_te^{i\Gamma_B}+Q^{*}_0b^{*}_te^{-i\Gamma_B}$. Therefore, the sidebands are linearly polarized. 
		
		If the laser fields are CWs, the statement in the previous paragraph is still valid for sidebands of all orders in the absence of inversion symmetry. We choose the same time-reversed two electron-hole recollision pathways as in the case of zero Berry connection, but with dipole vectors $\mathbf{d}_1[\mathbf{k}(t_0)]=a_{t_0}\sigma^{+}+b_{t_0}\sigma^{-}=\mathbf{d}^{*}_2[-\mathbf{k}(t_0)]$. Due to time-reversal symmetry, the dynamic phases and dephasing factors for the two pathways are the same, while the Berry phases are opposite. So from Eq.~\ref{apphsg_amplitude_abelian}, we have $\mathbf{P}_n\propto \mathbf{Q}_1-(-1)^n\mathbf{Q}_1^{*}$, where $\mathbf{Q}_1=e^{i\Gamma_B}Q_0(a_{t}\sigma^{+}+b_{t}\sigma^{-})$, i.e., $\mathbf{P}_n\propto \rho[e^{-i\varphi}\sigma^{+} + (-1)^ne^{i\varphi}\sigma^{-}]$. Thus, all sidebands are linearly polarized.
		
		In a band insulator with time-reversal and inversion symmetries, and in-plane dipole matrix elements being nonzero only between valence and conduction bands ($\mathbf{D}_{\text{inter}}=\mathbf{0}$), in general, an Abelian Berry connection can induce ellipticity from linear polarizations. Suppose, associated with each sideband, there are electron-hole pairs from different energy bands or there is more than one $k$-space trajectory. In this case, there is more than one pair of electron-hole recollision pathways related by time-reversal and inversion symmetries. As discussed above, each pair of recollision pathways contribute a linearly polarized amplitude to a sideband. In general, dynamic phases and Berry phases, obtained by electron-hole pairs from different energy bands or along different $k$-space trajectories, are not the same. The phase factor $\exp[i\Omega(t-t') + in\omega t]$ also depends on the $k$-space trajectory. Therefore, even if the Berry connection is Abelian, each sideband, as a sum of linear polarizations with different phases and polarization angles, can be elliptically polarized.

\subsection{Non-Abelian Berry connection}  \label{non_abelian_berry}		
		
		In a band insulator with time-reversal and inversion symmetries, even if there is only one $k$-space trajectory associated with each sideband, through inter-valence-band or inter-conduction-band transitions (which can be induced by a non-Abelian Berry connection, or nonzero inter-valence-band/inter-conduction-band transition dipole moments), time-reversed electron-hole pairs injected by a linearly polarized NIR laser can have nonzero total angular momentum at recollisions. We discuss the case of inter-valence-band transitions induced by a non-Abelian Berry connection. The discussion for the case of inter-valence-band/inter-conduction-band dipole transitions is similar. Consider the (100) GaAs QWs with only the lowest conduction subband and the highest two valence subbands as discussed in Section~\ref{semiclassical picture}. For simplicity, we neglect the dephasing effects, and further assume that the cellular functions for the valence subbands only involve $f_1\Ket{\frac{3}{2}, +\frac{3}{2}}$, $f_1\Ket{\frac{3}{2}, -\frac{1}{2}}$, $f_1\Ket{\frac{3}{2}, -\frac{3}{2}}$ and $f_1\Ket{\frac{3}{2}, +\frac{1}{2}}$. At a certain quasi-momentum $\mathbf{k}$, we can choose the cellular functions as related by time-reversal and inversion symmetries, so that the Berry connection matrices of the two spin sectors satisfy $\vec R_{\downarrow}=-\vec R_{\uparrow}^{*}$. When the QWs are resonantly excited by a linearly polarized NIR laser, the initial hole spinor states are $\phi_s=g_s(1,0)^T$ ($g_s$ is a constant with modulus 1). Angular momentum conservation law requires that the total angular momentum of the holes is zero initially. Right before inter-valence-band tunneling happens, the hole spinor states are still of the form $\phi_s=g'_s(1,0)^T$ ($g'_s$ is a constant containing the dynamic phase and Abelian Berry phase with modulus 1), and the total angular momentum of the holes remains zero, as discussed in the case of Abelian Berry connection. To see how non-Abelian Berry connection induce angular momentum changes of the holes, we calculate the spinor state $\phi_{\uparrow}$ from the dynamical equation, Eq.~\ref{second_eqn}, with initial condition $\phi_{\uparrow}(0)=g'_{\uparrow}(1,0)^{T}$ in two steps, considering only the off-diagonal elements of the Berry connection for the first step, and the dynamic phases and Abelian Berry phases for the second. In the first step, the hole spinor $\phi_{\uparrow}(0)$ evolves to $\phi_{\uparrow}(\Delta t)=g'_{\uparrow}(1,i\lambda\Delta t)^{T}$, where $\lambda=\dot{\mathbf{k}}(0)\cdot \vec{R}^{*}_{\uparrow,12}[\mathbf{k}(0)]$. Denote the cellular functions at $\mathbf{k}(\Delta t)$ as $\Ket{\text{HH}_{1,\uparrow}}_{\mathbf{k}(\Delta t)}=f_1(a_1\Ket{\frac{3}{2}, +\frac{3}{2}}+b_1\Ket{\frac{3}{2}, -\frac{1}{2}})$, and $\Ket{\text{HH}_{2,\downarrow}}_{\mathbf{k}(\Delta t)}=f_1(c_1\Ket{\frac{3}{2}, +\frac{3}{2}}+d_1\Ket{\frac{3}{2}, -\frac{1}{2}})$. The angular momentum of the hole spinor $\phi_{\uparrow} (\Delta t)$ can be calculated as
		\begin{equation}
		 J_{\uparrow}=(3/2)|a_1+i\lambda\Delta t c_1|^2-(1/2)|b_1+i\lambda\Delta t d_1|^2.
		 \end{equation}
Similarly, for $\phi_{\downarrow}(\Delta t)$ with initial state $\phi_{\downarrow}(0)=g'_{\downarrow}(1,0)^{T}$
 , we have		
 		\begin{equation}
		 J_{\downarrow}=(-3/2)|a_1^{*}-i\lambda^{*}\Delta t c_1^{*}|^2+(1/2)|b_1^{*}-i\lambda^{*}\Delta t d_1^{*}|^2.
		 \end{equation}
		 The total angular momentum of the holes is $J_{\uparrow}+J_{\downarrow}=0$. In the second step, the hole spinor $\phi_{\uparrow}(\Delta t)$ evolves into $\phi_{\uparrow}(2\Delta t)=g'_{\uparrow}(e^{i(\Gamma_{D,1}+\Gamma_{B,1})},e^{i(\Gamma_{D,2}+\Gamma_{B,2})}i\lambda\Delta t)^{T}$, where $\Gamma_{D,j}=-E_{cv,j}[\mathbf{k}(\Delta t)]{\Delta t}/{\hbar}$, $\Gamma_{B,j}=\vec{R}_{\uparrow, jj}[\mathbf{k}(\Delta t)] \cdot \dot{\mathbf{k}}(\Delta t)\Delta t$ ($j=1,2$) are the dynamic phase and Abelian Berry phase respectively. The hole spinor $\phi_{\uparrow}(\Delta t)$ get the same dynamic phases but opposite Berry phases, i.e., $\phi_{\downarrow}(2\Delta t)=g'_{\downarrow}(e^{i(\Gamma_{D,1}-\Gamma_{B,1}},e^{i(\Gamma_{D,2}-\Gamma_{B,2})}i\lambda\Delta t)^{T}$. Denote the cellular functions at $\mathbf{k}(2\Delta t)$ as $\Ket{\text{HH}_{1,\uparrow}}_{\mathbf{k}(2\Delta t)}=f_1(a_2\Ket{\frac{3}{2}, +\frac{3}{2}}+b_2\Ket{\frac{3}{2}, -\frac{1}{2}})$, and $\Ket{\text{HH}_{2,\downarrow}}_{\mathbf{k}(2\Delta t)}=f_1(c_2\Ket{\frac{3}{2}, +\frac{3}{2}}+d_2\Ket{\frac{3}{2}, -\frac{1}{2}})$. After some algebra, we can get the total angular momentum for the hole spinors at $2\Delta t$ as
		 \begin{equation}
		 J=2\Delta t \sin(\Delta\Gamma_D)\Re[(b_2^{*}d_2-3a_2^{*}c_2)\lambda^{*}e^{i\Delta\Gamma_B}],
		 \end{equation}
		 where $\Delta\Gamma_D=\Gamma_{D,2}-\Gamma_{D,1}$ and $\Delta\Gamma_B=\Gamma_{B,2}-\Gamma_{B,1}$. Thus, a nonzero total angular momentum is induced by electron-hole pairs associated with different energy bands.
		
\section{Semiclassical carrier dynamics} \label{appSemiclassical}
		\begin{figure*}
			\includegraphics{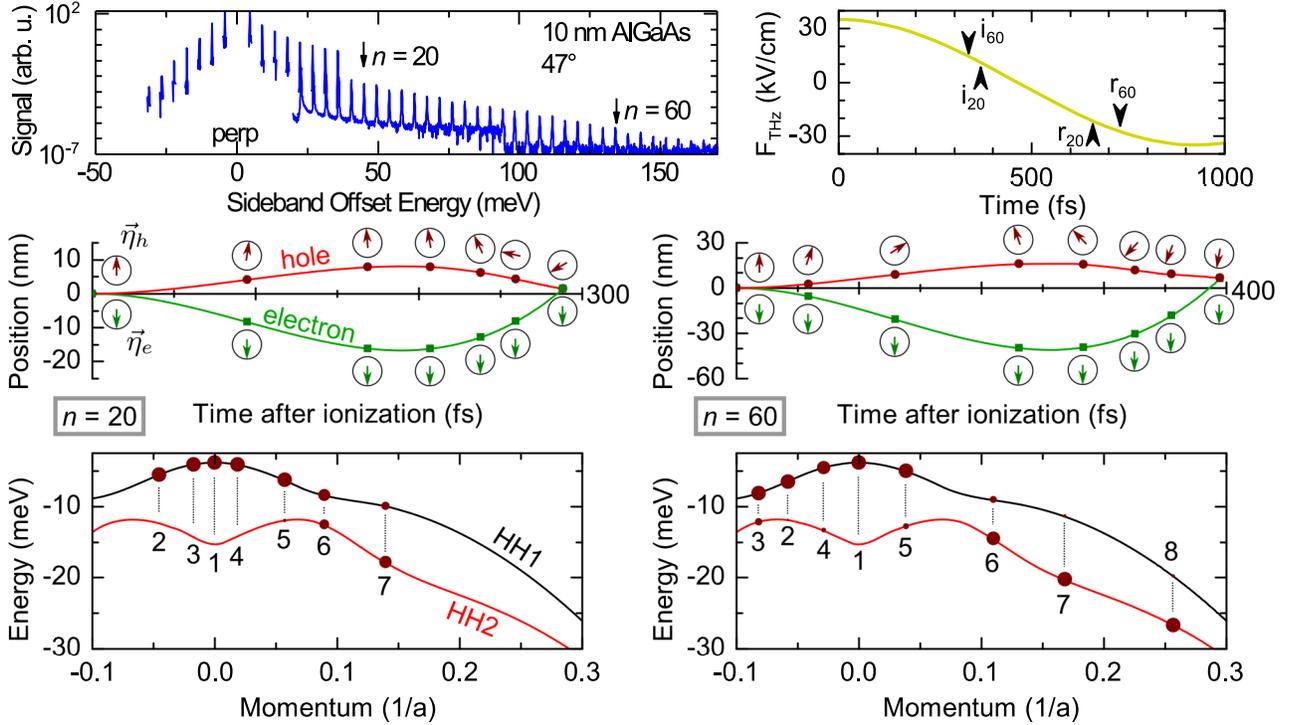}
			\caption{Semiclassical trajectories for the $n = 20$ and $n = 60$ sidebands in the 10 nm AlGaAs sample. (upper left) Full perpendicular HSG spectrum experimentally measured in the 10 nm AlGaAs sample oriented at 47$^\circ$. (upper right) Time trace of the THz electric field.  The arrows point to the time instants of ionization, $i_n$, and recollision, $r_n$, for the two sidebands considered here. (middle left) The real-space trajectories of the electron and hole for the 20th order sideband. The spin state is $\vec{\eta}_e = [\uparrow, \downarrow]^T$ for the electron, and $\vec{\eta}_h = [\ket{\text{HH1}}, \ket{\text{HH2}}]^T$ for the hole. Each arrow represents a spin/pseudo-spin in xz-plane for each of seven instants throughout the trajectory. (lower left) The location of the hole in the valence subbands at each of those seven instants with the area of each maroon circle representing the relative weight in either subband. (middle right) The real-space trajectories of the electron and hole for the 60th order sideband, with the spinor directions drawn for eight instants. The hole spinor almost entirely flips to the HH2 state at recollision. (bottom right) The location of the hole in the valence subbands at each of the eight instants. For both cases, the electron is always in spin-down state and the y-component of the hole pseudo-spin is approximately zero. Comparing the two cases shows the scale of the effects of non-Abelian Berry curvature.}
			\label{appFigSemiclassical}
		\end{figure*}
		To demonstrate how the non-Abelian Berry connection affects the dynamics of the electron-hole pairs, two semiclassical trajectories with different relative phases between the NIR and THz laser fields are shown in Fig.~\ref{appFigSemiclassical}. As can be seen in Fig.~\ref{pol_comp}, the relative strength of the parallel and perpendicular sidebands of 20th and 60th orders are very different for the 10 nm AlGaAs QWs. The phases of the THz field for ionization and recollision for the two orders of HSG are labeled in the upper right of Fig.~\ref{appFigSemiclassical}. The trajectory associated with the 60th order sideband is about 100 fs longer than the trajectory associated with the 20th order sideband.

		The details of the trajectory associated with the 20th order sideband are plotted in the lower left two graphs in Fig.~\ref{appFigSemiclassical}. Over the course of the acceleration step, the electron and hole paths separate by almost 30 nm at their farthest. The spinor states of the two particles are plotted at seven different time instants (the spinor directions are chosen for excitation by a $\sigma^{+}$ NIR photon). The spin state of the electron does not change, but the hole pseudo-spin rotates by a large amount in the last 50 fs. Notice that the position at recollision is not at exactly zero, but slightly positive. This translation results mainly from the non-parabolic nature of the hole subbands. If the masses of the electron and hole remained the same in the entire process, the recollision would occur at exactly zero. The hole states at the same seven instants are shown in the band structure underneath, where the area of each circle represents the probability amplitude of being in the subband. At instant $\#1$, the hole is entirely in the HH1 subband. As the hole accelerates to the left, at instant $\#2$, it only slightly mixes with the HH2 subband. As the hole accelerates to the right, at instants $\#3-7$, the hole spinor rotates significantly as it passes through the avoided crossing point of HH1 and HH2 subbands. The majority of the spinor weight is then in the HH2 subband.

		The details of the 60th order trajectory are plotted in the lower right two graphs in Fig.~\ref{appFigSemiclassical}. The non-Abelian Berry curvature has a much stronger effect. The electron and hole travel much further apart for this trajectory, and the hole spinor rotates more substantially, shown now at eight instants in the trajectory. The location of recollision is shifted almost 10 nm away from the origin. At $t=120$ fs, instant $\#3$, the hole sits close to the avoided crossing point, and some noticeable weight is transferred to the HH2 subband. As it accelerates to the right, at instant $\#4$, some of the weight is transferred back to the HH1 subband. As it continues accelerating to the right in the last 100 fs, the hole almost completely tunnels to the HH2 subband. 

\section{Effects of well width}\label{width_Berry}
\begin{figure}
			\includegraphics[width=0.5\textwidth]{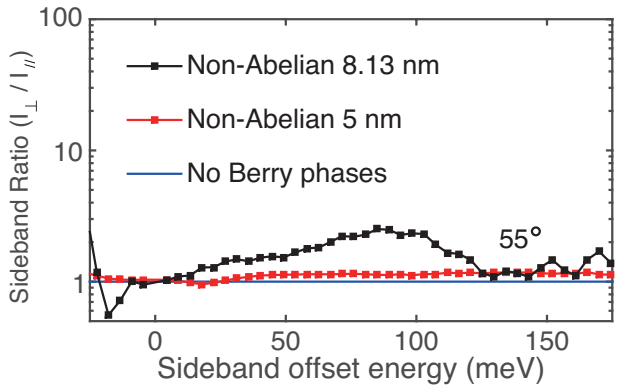}
			\caption{Theoretical calculation of sideband intensity ratio $I_{\bot}/I_{\parallel}$ for the 5 nm GaAs sample at $55^{\circ}$ with two different well widths for the conduction band, 5 nm (red curve) and 8.13 nm (black curve replotted from Fig.~\ref{expt_theory}). The blue line is the result if the Berry connection is assumed to be zero.}
			\label{appFig_width_berry}
		\end{figure}

       We show that the deviation of the ratio $I_{\bot}/I_{\parallel}$ between experiment and theory for the 5 nm GaAs QWs at $55^{\circ}$ could be explained by the overestimation of Berry connection in the calculation. In Fig.~\ref{expt_theory}, the effective well width for the conduction band is taken to be $L_e=8.13$ nm in the calculation. As shown in Fig.~\ref{appFig_width_berry}, if we use $L_e=5$ nm without changing all other parameters, we obtain a sideband ratio close to 1, as observed in the experiment. 
        The conduction subbands have higher energy levels for narrower QWs. If the effective mass of the conduction subbands do not depend on the effective well width $L_e$, as in our calculation, the $k$-space region, in which the conduction band energy of the QWs lies below that of the AlGaAs barriers, will be smaller for narrower QWs. For QWs with $L_e=5$ nm, at $55^{\circ}$, the largest momentum that an electron can have before it steps into the barrier is calculated to be about 0.08 1/a. To generate a sideband, the electron and hole should have the same momentum, so the relevant $k$-space region in HSG for the valence subbands lies between $\pm0.08$ 1/a along the lattice direction at $55^{\circ}$. In this $k$-space region, the highest valence subband is nearly parabolic (see Fig.~\ref{orientation}) and the Berry curvature is close to zero, which implies that there is almost no dynamical birefringence. For QWs with $L_e=8.13$ nm, a hole has chances to pass the avoided crossing point, where the Berry curvature is relatively large, which could induce a larger degree of dynamical birefringence. Therefore, we might have had an overestimation of the Berry connection by using a larger effective well width in the calculation in Fig.~\ref{expt_theory}.

\section{Dynamical Jones Calculus} \label{DynamicalJonesCalculus}
    The Jones calculus is a convenient formalism for describing the propagation of perfectly (or fully) polarized light through linear optical media and components~\cite{jones1941new}. The Jones calculus manipulates the Jones vector, a complex two-component vector that can only describe perfectly polarized light.  This formalism handles interference phenomena naturally and can be used with any orthogonal polarization state basis, such as linear or circular. Because the NIR laser and sidebands are perfectly polarized, as was shown in Sect.~\ref{Comparison}, and interference is central to our model, we generalize the Jones calculus to the nonlinear optical phenomenon of HSG.

	Familiar linear optical elements, like polarizers and wave plates, can each be associated with a conventional Jones matrix. In the dynamical Jones calculus, the THz-driven quantum well acts as the optical element, and the Jones matrix relates the polarization state of each sideband to that of the incident NIR laser. For each sideband, we assign a Jones matrix $\mathcal{J}_n=J_{ij,n}$, which is defined as
		\begin{equation}
			\begin{pmatrix}
				E_{x,\text{HSG}}\\
				E_{y,\text{HSG}}
			\end{pmatrix}_{\!n} =
			\begin{pmatrix}
				J_{xx} & J_{xy}\\
				J_{yx} & J_{yy}
			\end{pmatrix}_{\!n}
			\begin{pmatrix}
				E_{x,\text{NIR}}\\
				E_{y,\text{NIR}}
			\end{pmatrix}.\label{appEqJones}
		\end{equation}
    Following the input of a NIR laser field described by Jones vector $(E_{x,\text{NIR}},E_{y,\text{NIR}})^{T}$, the THz-driven quantum well produces HSG with Jones vector $(E_{x,\text{HSG}},E_{y,\text{HSG}})^{T}$. The elements of dynamical and conventional Jones vectors and Jones matrices are in general complex. If HSG were an isotropic effect, $\mathcal{J}_n$ would be proportional to the identity matrix. Dynamical linear birefringence is the dynamical analog to the familiar linear birefringence that is observed in a material like calcite. In the case of pure dynamical linear birefringence, $\mathcal{J}_n$ is diagonal, with diagonal elements having different complex phases.
    
    The $\mathcal{J}_n$ can be determined experimentally. Measurements of the polarization angles $\alpha$ and $\gamma$ for more than three different polarization states of the NIR laser determine ${J_{xy,n}}/{J_{xx,n}}$, ${J_{yx,n}}/{J_{xx,n}}$, ${J_{yy,n}}/{J_{xx,n}}$. Together with one measurement of sideband intensity for a certain NIR laser field, the dynamical Jones matrix can be determined to within an overall phase factor, which can be measured through time-resolved experiments (see Supplementary Material~\cite{Note12} for more details).
    
    In the experiment, the polarization angles of the sidebands and their intensities are measured independently. However, both the polarization state and the intensity of a sideband are determined by the dynamical Jones matrix $\mathcal{J}_n$. Using the $\mathcal{J}_n$ formulation, the sideband intensity ratio $I_\bot / I_\parallel$ can be derived from the polarization ellipse measurements. This provides a way to check the consistency of the experiments.
    
    \begin{figure}
    	\includegraphics{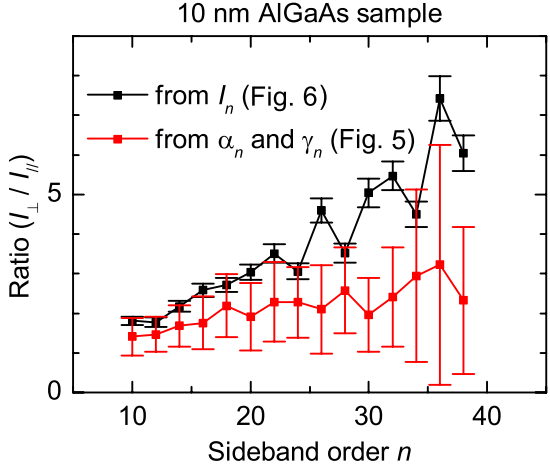}
    	\caption{Dynamical birefringence ratio $I_\bot / I_\parallel$ from two independent measurements.  The black squares are from the intensity measurements (center left, Fig.~\ref{expt_theory}). The red squares are obtained from the unitary Jones matrices $\mathcal{J}_n$ calculated from the polarization state angle measurements from Fig.~\ref{pol_state}.}
    	\label{appFigJonesRatio}
    \end{figure}
    
    Fig.~\ref{appFigJonesRatio} compares the ratio $I_{\bot}/I_{\parallel}$ calculated from the polarization state measurements (from Fig.~\ref{pol_state}(b)) with the one calculated from the sideband intensity measurements (from center left of Fig.~\ref{expt_theory}). To calculate the ratio from the polarization states, Eq.~\ref{appEqJones} was used with the three data sets from Fig.~\ref{pol_state}(b) to calculate ${J_{xy,n}}/{J_{xx,n}}$, ${J_{yx,n}}/{J_{xx,n}}$, and ${J_{yy,n}}/{J_{xx,n}}$. The values were then used in Eq.~\ref{appEqJones} to calculate the expected sideband intensity ratio given an input NIR laser polarization state. It should be noted that no information of the intensities of the sidebands is used, only the polarization states of the sidebands and NIR laser. 
    
    The polarization state measurements remarkably well with the intensity measurements, matching the trend of increasing sideband intensity ratio with increasing order, as well as the overall scale. Agreement between techniques is not as good for the GaAs sample (see Supplementary Material~\cite{Note2}). The deviation between the two methods is likely due to a systematic error in the Stokes polarimeter used to measure the NIR laser and sideband polarization states. The polarimeter is sensitive to the exact retardance value of the quarter wave plate used (see Supplementary Material~\cite{Note2} for details). Improving the accuracy of these measurements is outside the scope of the current work. The internal consistency between the sideband intensity measurements and the polarization state measurements---which is independent of any inputs to simulation---strongly supports the validity of our theoretical approach.
    
    Jones matrices can also be computed from the theory and compared with experiment. In the context of theory, it is natural to express the Jones vectors and matrices on the basis of circularly polarized states $\sigma^{\pm}$. With this basis, we define the Jones matrices $\mathcal{T}_n = T_{ij,n}$ as
    	\begin{equation}
    		\begin{pmatrix}
    			\sigma^{+}_\text{HSG}\\
    			\sigma^{-}_\text{HSG}
    		\end{pmatrix}_{\!n} =
    		\begin{pmatrix}
    			T_{++} & T_{+-}\\
    			T_{-+} & T_{--}
    		\end{pmatrix}_{\!n}
    		\begin{pmatrix}
    			\sigma^{+}_\text{NIR}\\
    			\sigma^{-}_\text{NIR}
    		\end{pmatrix}.
    	\end{equation}
    Following the input of a NIR laser field described by Jones vector $(\sigma^{+}_\text{NIR}, \sigma^{-}_\text{NIR})^T$, the THz-driven quantum well produces HSG with Jones vector $(\sigma^{+}_\text{HSG}, \sigma^{-}_\text{HSG})^T$. The Jones matrices $\mathcal{T}_n$ and $\mathcal{J}_n$ are related by a unitary transformation. If HSG were an isotropic effect, $\mathcal{T}_n$ would again be proportional to the identity matrix. Dynamical circular birefringence is analogous to the familiar circular birefringence--also called optical activity--that is observed, for example, when the polarization of visible light is rotated upon transmission through a solution containing a chiral molecule like glucose. In the case of pure dynamical circular birefringence, $\mathcal{T}_n$ is diagonal, with diagonal elements having different complex phases. In general, HSG results in both dynamical linear and circular birefringence, so all of the matrix elements of dynamical Jones matrices, in either linear or circularly polarized bases, are non-zero and complex.

%

\end{document}